\documentclass[journal]{IEEEtran}
\ifCLASSINFOpdf
\else
\fi
\usepackage{color}
\usepackage{enumerate}
\usepackage[cmex10]{amsmath} 
\usepackage{siunitx}
\usepackage{import} 
\usepackage{enumitem}
\usepackage{indentfirst} 
\usepackage{standalone}
\usepackage[section]{placeins} 
\usepackage{pdflscape}
\usepackage{makecell}
\usepackage{caption}
\usepackage{tabularx}
\usepackage{algorithm} 
\usepackage{amssymb} 
\usepackage{microtype}
\usepackage{soul} 
\usepackage{mathtools}
\usepackage{multirow}
\usepackage{verbatim}
\usepackage{graphicx}
\usepackage{pgfplots}

\usepackage{subfloat} 
\usepackage{verbatim}
\usepackage[utf8]{inputenc}
\usepackage[T1]{fontenc}
\usepackage{amsmath} 
\usepackage[shellescape,latex]{gmp} 
\usepackage[space]{grffile}
\usepackage{booktabs} 
\usepackage{xfrac}
\usepackage{tabularx}
\usepackage{cite}
\usepackage{adjustbox}
\usepackage{verbatim}
\usepackage{url}
\usetikzlibrary{shapes}
\usepackage{multicol, blindtext}    
\usepackage[]{color}
\usepackage{framed}
\usepackage{alltt}
\usepackage{bbm}
\usepackage[T1]{fontenc}
\usepackage{array}
\usepackage{colortbl}
\usepackage{booktabs}
\usepackage{multirow}
\usepackage{eurosym}
\usepackage{listings}
\usepackage{ifthen}
\usepackage{pgfplots}
\usepackage{pgfplotstable}
\usepackage{pgfcalendar}
\usepackage{pgfgantt}
\usepackage{tikz}
\usetikzlibrary{shapes.geometric, arrows}
\usepackage{longtable}
\usepackage{amsfonts}
\usepackage{subfigure}
\usepackage{float}
\usepackage{afterpage}
\usepackage{mathrsfs}
\usepackage{algpseudocode} 
\usepackage{textcomp}
\usepackage{xspace}
\makeatletter
\newcommand\fs@betterruled{
  \def\@fs@cfont{\bfseries}\let\@fs@capt\floatc@ruled
  \def\@fs@pre{\vspace*{6pt}\hrule height.8pt depth0pt \kern2pt}%
  \def\@fs@post{\kern2pt\hrule\relax}%
  \def\@fs@mid{\kern2pt\hrule\kern2pt}%
  \let\@fs@iftopcapt\iftrue}
\floatstyle{betterruled}
\restylefloat{algorithm}
\makeatother
\DeclareCaptionFont{eightpt}{\fontsize{8pt}{9pt}\selectfont}
\usepackage{fmtcount}

\begin{document}
\title{
Multi-Agentic AI for Fairness-Aware and Accelerated Multi-modal Large Model Inference in Real-world Mobile Edge Networks
}

\author{Haiyuan Li, Hari Madhukumar, Shuangyi Yan, Yulei Wu, Dimitra Simeonidou

\vspace{-0.2cm}

\thanks{This work was supported in part by the UK-funded project REASON under the Future Open Networks Research Challenge sponsored by the Department of Science, Innovation and Technology (DSIT), and in part by the NVIDIA Academic Grant Program, the UK Engineering and Physical Sciences Research Council (EPSRC) grant EP/Y037243/1, EP/X04047X/2 for TITAN and the Federated Telecoms Hub, and grant EP/Y036514/1 for JOINER project.}
 
\thanks{H. Li, H. Madhukumar, S Yan, Y. Wu, D. Simeonidou are with the Smart Internet Lab, Department of Electrical and Electronic Engineering, University of Bristol, BS8 1QU, U.K. (e-mail: ocean.h.li@bristol.ac.uk).}%

\thanks{Manuscript received XXXX YY, ZZZZ; revised XXXX YY, ZZZZ.}
}

\markboth{\LaTeX\ Class Files,~Vol.~X, No.~Y, Month~202X}%
{Shell \MakeLowercase{\textit{et al.}}: Bare Demo of IEEEtran.cls for Journals}

\maketitle 
\begin{abstract}
Generative AI (GenAI) has transformed applications in natural language processing and content creation, yet centralized inference remains hindered by high latency, limited customizability, and privacy concerns. Deploying large models (LMs) in mobile edge networks emerges as a promising solution. However, it also poses new challenges, including heterogeneous multi-modal LMs with diverse resource demands and inference speeds, varied prompt/output modalities that complicate orchestration, and resource-limited infrastructure ill-suited for concurrent LM execution. 
In response, we propose a Multi-Agentic AI framework for latency- and fairness-aware multi-modal LM inference in mobile edge networks. Our solution includes a long-term Global Planning Agent, a short-term Prompt Scheduling Agent, and multiple on-node LM Deployment Agents, all powered by foundation language models.  These agents cooperatively optimize prompt routing and LM deployment through natural language reasoning over runtime telemetry and historical experience. 
To evaluate its performance, we further develop a city-wide testbed that supports network monitoring, containerized LM deployment, intra-server resource management, and inter-server communications.
Experiments demonstrate that our solution reduces average end-to-end latency by 42.9\% compared with a state-of-the-art online-learning baseline and by 80.6\% compared with the full-activation baseline, while maintaining a mean service-type Jain's index of 0.925. Moreover, our solution adapts quickly without fine-tuning, offering a generalizable solution for optimizing GenAI services in edge environments.
\end{abstract}

\section{Introduction}
\label{sec:introduction}
Generative AI (GenAI) is reshaping modern technology, powered by multi-modal large models (LMs) built on the Transformer framework. Utilizing embeddings, self-attention, and multi-layer perceptrons~\cite{vaswani2017attention, peebles2023scalable}, these models capture complex patterns in text, images, and videos, enabling tasks such as human-like conversations, document summarization, realistic image generation, and video analysis. Generative AI-as-a-Service is evolving into a pivotal role in next-generation networks~\cite{xu2024unleashing, wang2023overview}. However, current GenAI service delivery heavily depends on centralized data centers, facing significant challenges such as uplink bandwidth costs, transmission latency, privacy risks, and limited model customization. In response, mobile edge computing (MEC) offers a promising alternative infrastructure by shifting inference tasks from data centers to edge servers~\cite{chen2024netgpt, lin2025pushing}, allowing LMs to be distributed across MEC servers, balancing computational efficiency, reducing transmission latency, and eliminating privacy concerns. 

To fully exploit the potential of edge LM inference, effective task scheduling and resource management are essential for improving system performance and service quality.
Numerous studies have explored optimization in edge computing, with computational tasks in these works generally falling into two categories. The first involves general-purpose services~\cite{fan2023game, liu2023joint, yang2023cooperative, li2022drl, xiao2022multi, ren2022efficient, cang2023online}, where tasks are defined in domains such as autonomous driving, IoT, or semantic extraction, typically by simulating network resource requirements, such as bandwidth and central processing unit (CPU) usage. As AI inference demand increases, a second category has emerged that focuses on edge AI applications~\cite{gao2025optimizing, aghapour2023task, li2019edge, zhou2021bbnet, dai2021cinet, hu2019dynamic, eshratifar2019jointdnn, jeong2018ionn}, where computing tasks typically involve executing specific neural network models on associated datasets. 
While these studies offer significant benefits, 
LM inference introduces several unique and complex characteristics that distinguish it from conventional tasks. These characteristics lead to a range of challenges, including the following key examples:

\noindent \textit{i) Excessive resource demands \& simulation fidelity gap:} Unlike traditional computing tasks, the substantial size of LMs necessitates extensive storage and memory, often leading to out-of-memory errors and limiting model parallelization. Moreover, system-level support (e.g., specialized software packages, operating systems, and plugins) and hardware heterogeneity across mobile edge networks further complicate the task deployments. These constraints lie beyond the scope of simulations, as they often involve complex, dynamic factors and dependencies that are difficult to model in synthetic environments. \\
\noindent \textit{ii) LM heterogeneity and concurrency:} Multi-modal LMs exhibit inherent performance and demand heterogeneity. For instance, the inference speed of image-based models is often slower compared to text-based models. Additionally, their resource demands vary substantially; some LMs necessitate extensive video random access memory (vRAM) and RAM, while others prioritize graphics processing unit (GPU) usage with less (v)RAM requirements. This heterogeneity presents a unique challenge in balancing the concurrent operation of diverse LMs within a shared resource-constrained environment.\\
\noindent \textit{iii) Traffic diversity:} Unlike traditional tasks, the LM input/output (i.e., prompts/inference results) exhibit high diversity. These prompts may include text generation, image generation, image interpretation, and more. During collaborative edge operations, such diverse prompts can introduce varying levels of transmission overhead during both upstream prompt submission and downstream inference result delivery.\\
\noindent \textit{iv) LM-specific runtime uncertainty:} Different model modalities exhibit different cold-start times, memory footprints, GPU sensitivities, and inference-latency characteristics. In addition, the same LM can also exhibit different service rates under different GPUs, runtime libraries, container images, and concurrent workloads. This variability makes it impractical to rely on a single fixed performance profile per LM, as any profile derived under one deployment context may no longer hold when the hardware configuration, software stack, or co-located workloads differ from the original profiling conditions.

Taken together, these challenges reveal a structural tension in balancing diverse and latency-sensitive user intents with highly heterogeneous model demands, all under constrained and fragmented resource conditions. 
While some efforts have been made to optimize the computation offloading of LMs~\cite{zhang2024edgeshard, zhang2024edge, chen2024adaptive, li2024multi, fang2023large, du2024diffusion, xu2023joint, yuan2024generative, yang2024perllm, xuchen, liang2025networked}, these studies have not fully addressed these challenges, and their limitations can be summarized as follows:

\begin{itemize}[leftmargin=*]
    \item Current strategies predominantly focus on server selection for large language model (LLM) execution, while overlooking the dynamic and fine-grained allocation of resources across multi-modal LMs. This limitation ignores model heterogeneity and fairness in scheduling and resource allocation for concurrent tasks with diverse requirements.
    \item Although prompt allocation across edge servers has been explored in prior work, the transmission latency associated with transferring prompts and inference results has been largely overlooked. This oversight impacts total inference time and degrades the effectiveness of scheduling and resource allocation strategies.
    \item In practice, foundation models and task-specific LMs are frequently introduced, updated, or deprecated. Classic AI-driven resource management schemes, which rely on extensive training data and lack generalization capability, often struggle to adapt to the non-stationary and evolving nature of edge LM deployments.
    \item Most existing approaches rely on overly simplified simulations, undermining the feasibility of their strategies. Key factors such as the LM initialization duration, the disruption of timely feedback in decision-making processes caused by LM runtime delays, the nonlinear relationship between resource usage and computation speed, the overhead of control signaling, and the resumable execution of interrupted LM service are often overlooked.
\end{itemize}

To bridge these critical gaps, the main contributions of this paper are summarized as follows:

\begin{itemize}[leftmargin=*]
    \item \textit{Multi-objective optimization for multi-modal LM inference:} We formulate a multi-objective optimization problem for managing multi-modal LM inference in mobile edge networks, where diverse service types, such as text-to-text, text-to-image, and image-to-text, compete for limited computational resources. The proposed formulation jointly optimizes two performance criteria, namely inter-LM fairness and end-to-end latency, while respecting network resource constraints.
    \item \textit{GenAI for networks \& networks for GenAI:} We propose a two-tiered Agentic AI framework that integrates LLM-based agents across both long-term planning and short-term decision-making layers for optimizing edge LM inference. The Tier-1 Global Planning Agent operates at a slower temporal resolution, processing compact summaries of long-term telemetry and request distribution. It retrieves historical cases from episodic memory and synthesizes a macro-level policy that specifies probabilistic prompt-to-server routing and node-level LM deployment intents. At a finer temporal granularity, a Prompt Scheduling Agent translates this strategy into concrete routing decisions by monitoring real-time cluster conditions, while distributed LM Deployment Agents dynamically adjust local LM activations and resource allocations.   
    \item \textit{Practical deployment:} The proposed framework models the full lifecycle of prompt execution, including reception, scheduling, queuing, inference, and result delivery, capturing the end-to-end operational flow for realistic optimization. To enable practical validation, we develop a city-wide edge testbed based on OpenStack and Kubernetes, equipped with real-time monitoring, inter-server communication, and intra-server resource adjustment capabilities for LM deployment. 
    \item \textit{Extensive experiment:} The proposed solution significantly reduces end-to-end latency while improving service success ratios, achieving a well-balanced trade-off between responsiveness and fairness. Relative to the full-activation baseline, our system reduces average end-to-end latency by 80.6\% while maintaining a mean Jain's fairness index of 0.925 across LM service types. Against PerLLM and MAL, the proposed multi-agentic framework reduces latency by 42.9\% and 21.4\%, respectively, demonstrating a better latency--fairness trade-off under delayed and non-stationary feedback. Moreover, our framework demonstrates strong generalization ability through episodic memory and few-shot learning. By leveraging the reasoning capability of LLM, the system rapidly adapts to changing workload patterns without requiring task-specific fine-tuning. Together, these findings highlight the practicality and scalability of LLM-driven agentic AI for autonomous GenAI service management in mobile edge networks.
\end{itemize}

The remainder of this paper is constructed as follows. 
Section~\ref{sec:literature} presents related work. 
Section~\ref{sec:problem} describes the Edge LM operation scenario and formulates the optimization problem. 
Section~\ref{sec:Methodology} discusses the technical details of the proposed solution.
Then, Section~\ref{sec:testbed} presents the setup of the testbed, deployment of various LMs, and implementation of the solution.
The experimental results are provided in Section~\ref{sec:results}.
Finally, the paper concludes with a summary of our key findings in Section~\ref{sec:conclusion}.

\vspace{-0.3cm}
\section{Literature Review}
\label{sec:literature}

This paper focuses on optimizing LM operation at the edge.
To enable efficient edge deployment of LMs, existing research primarily falls into two categories: (i) 
\textit{model-centric approaches}, which reduce the model size through techniques such as quantization and weight pruning~\cite{yu2024edge, lin2024awq, ma2023llm}; and (ii) \textit{resource-centric approaches}, which optimize the resource allocation to improve the LM execution efficiency~\cite{zhang2024edgeshard, zhang2024edge, chen2024adaptive, li2024multi, fang2023large, xu2023joint, du2024diffusion, yuan2024generative, yang2024perllm, xuchen, liang2025networked}. 

\subsubsection{Model-centric approaches}
To reduce LM size in fitting edge devices, recent state-of-the-art (SOTA) approaches have converged on hybrid pipelines, systematically combining pruning, low-bit quantization, and low-rank factorization methods. For instance, Yu et al.~\cite{yu2024edge} introduced a layer-wise unified compression approach. This method dynamically adjusts quantization and pruning based on the sensitivity of each layer to compression, thereby reducing the computational load. In addition, the authors proposed an adaptive layer-tuning mechanism and a voting strategy to further minimize memory usage. This is achieved by selectively updating specific layers during inference and by implementing a voting mechanism to select the optimal output. Similarly, based on the same principle that not all model weights are equally important for performance, Lin et al.~\cite{lin2024awq} proposed an activation-aware weight quantization technology. This approach mitigates quantization errors without the need for retraining by selectively preserving the most critical 1\% of weights, identified through activation magnitudes, and applying per-channel scaling.
Dropout has also been used as a partial-model training mechanism to reduce communication and computation overhead in federated edge learning. Wen et al.~\cite{wen2022federated} apply this idea by sampling reduced subnets from the global model and assigning them to resource-constrained devices for local training, with heterogeneous dropout rates adapted to device/channel conditions. This reduces both model-transmission overhead and local update computation compared with full-model federated learning. Xie et al.~\cite{xie2026fedlodrop} extend dropout-based partial training to federated LoRA fine-tuning of LLMs by applying dropout to the trainable low-rank matrices, while jointly considering generalization behavior and resource constraints such as latency and device energy consumption.
These compression and dropout-based techniques have significantly improved the feasibility of deploying LLMs on resource-constrained edge devices. However, efficient runtime execution remains challenging due to the limited and heterogeneous nature of edge computing environments.

\subsubsection{Resource-centric approaches}
With the model optimized for edge deployment, this paper focuses on efficient LLM inference through network resource coordination. Within the existing solutions, the LMs may be deployed either in their entirety or as split components, with the allocated resources potentially residing on edge devices, MEC servers, or the cloud. In specific, Du et al.~\cite{du2024diffusion} designed an actor-critic-based deep reinforcement learning (DRL) solution where the actor is replaced with a diffusion model capable of generating discrete decisions. This model selects the most appropriate AI content service providers, considering the available resources of each service provider. However, this approach fails to consider the internal resource management within each service provider, as it focuses on the execution destination of the LMs while neglecting the available resources allocated for running the model.
In contrast, with the objective of minimizing inference latency and maximizing prediction accuracy, Yang et al.~\cite{fang2023large} propose a rewardless DRL-based algorithm to address the generalization limitations observed in reward-dependent models in dynamic environments. Specifically, the algorithm selects the optimal offloading destination between an edge server and a data center (instead of AI service providers), dynamically allocating computational and network resources based on the remaining available resources. However, this solution overlooks inter-edge-server collaboration, which may lead to suboptimal resource utilization, restricted service coverage, and imbalanced system performance. 
In response, Yang et al.~\cite{yang2024perllm} optimized the resource allocation between servers by formulating the problem as a multi-armed bandit (MAB) and addressing it using an upper confidence bound-based algorithm. This approach enables the selection of the most suitable server to execute the LLM for each service. However, their solution assumes the availability of a single language model, whereas in practice, a variety of multi-modal LMs often coexist within the network.
Yuan et al.~\cite{yuan2024generative} applied a similar solution by also formulating the problem as an MAB with the objective of minimizing transmission and inference energy. In contrast, they took it a step further by not only selecting the most appropriate MEC server for the requests but also choosing the most suitable model from five available LLMs based on service needs. Nevertheless, despite offering multiple LLM options, their approach overlooks the competition for memory and computing resources when running multiple LLMs simultaneously. Additionally, it fails to account for request queuing delays, as not every request can be processed immediately after the allocation decision.

Beyond DRL- and MAB-based resource allocation, recent studies have begun to explore agentic AI and heterogeneous runtime mechanisms for managing edge AI applications. Dehury et al.~\cite{dehury2026llm} propose an LLM-assisted edge intelligence framework in which cloud-hosted LLMs generate and iteratively regenerate lightweight execution logic, which is then validated on edge devices before deployment. In smart-space systems, Saleh et al.~\cite{saleh2025usercentrix} introduce a memory-augmented agentic framework that combines intent reasoning, urgency-aware decision control, and memory-based adaptation to balance latency, accuracy, and computational cost. Along this memory-centric direction, Saleh et al.~\cite{saleh2025memindex} further develop MemIndex, which provides agentic event-based distributed memory management for LM-based multi-agent pub/sub systems, using intent-indexed structures and agent negotiation to support storage, retrieval, update, deletion, and offloading operations. These studies indicate a broader shift from static edge offloading toward adaptive AI-system operation, including agentic coordination, memory-aware management, and runtime adaptation. This paper addresses the complementary problem of runtime orchestration for multi-modal LM services across MEC servers, where prompt routing, queueing, and LM deployment are jointly coupled.

\vspace{-0.2cm}
\noindent \section{System Model and Problem Formulation}
\label{sec:problem}

\begin{figure}[t]
    \centering
    \setlength{\abovecaptionskip}{0.2cm}
    \includegraphics[width=1\linewidth]{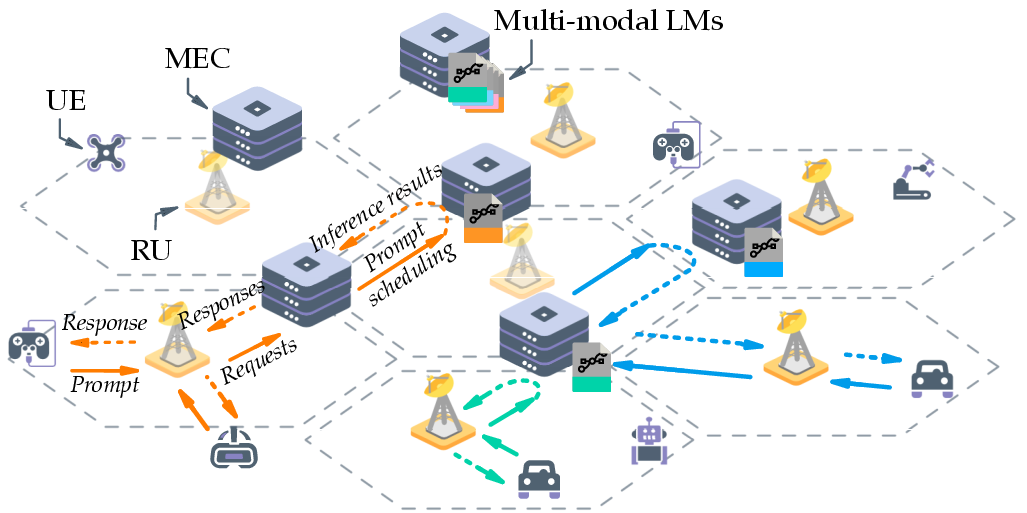}
    \caption{A mobile edge network example where MEC servers host LMs of different modalities and collaboratively handle diverse inference requests.}
    \label{fig:Orchs}
\end{figure}

We consider a multi-server mobile edge network, as illustrated in Figure~\ref{fig:Orchs}. The network consists of user equipment (UEs), radio units (RUs), and a set of heterogeneous MEC servers with different resource capacities. In this network, UEs generate service requests for various LM-based tasks, such as text-to-text processing, text-to-image generation, and image-to-text summarization. These requests are transmitted to the corresponding RUs over wireless channels. Each RU aggregates the received prompts within its coverage area and forwards them to its associated MEC server~\cite{chen2015efficient}.

Let $\mathcal{N} = \{1, 2, \ldots, N\}$ denote the set of MEC servers, and $\mathcal{T} = \{1, 2, \ldots, T\}$ represent the discrete time slots over an arbitrarily large time duration. 
At each time slot $t \in \mathcal{T}$, each MEC server $n \in \mathcal{N}$ receives a set of LM service requests, denoted by $\mathcal{Q}_{nt} = \{ q_i \}_{i \in \mathcal{I}_{nt}}$, where $\mathcal{I}_{nt} \subseteq \mathcal{I}$ is the subset of LM types that received prompts at server $n$ during slot $t$. 
Each request $q_i$ corresponds to a batch of $K_q$ prompts of type $i$. The prompts may be text or images, depending on the modality and functionality of the associated LM.
As illustrated in Figure~\ref{fig:Orchs}, prompts of request $q \in \mathcal{Q}_{nt}$ can be transferred among MEC servers to locate one hosting the required LM of type $i(q)$ for inference. Each LM $i \in \mathcal{I}$ requires a minimum amount of RAM and GPU memory, denoted by $M_i^{\min}$ and $V_i^{\min}$, respectively. Notably, there are no hard requirements on the number of CPU or GPU cores for LM deployment. However, the amounts of GPU and CPU resources allocated to an active model, denoted by $G_{ni}$ and $C_{ni}$, directly influence its inference speed once deployed.
Based on the received request, each server $n \in \mathcal{N}$ can select a subset of LMs $\mathcal{I}_n^{\text{act}} \subseteq \mathcal{I}$ to activate, depending on its available resources. 
The following constraints ensure that the total consumption of resources by all active LMs does not exceed the physical limits of each server:
\begin{equation}
    \begin{aligned}
    &\sum_{i \in \mathcal{I}_n^{\text{act}}} G_{ni} \leq G_n, \ \ 
    \sum_{i \in \mathcal{I}_n^{\text{act}}} C_{ni} \leq C_n, \\
    &\sum_{i \in \mathcal{I}_n^{\text{act}}} M_i^{\min} \leq M_n,\ \
    \sum_{i \in \mathcal{I}_n^{\text{act}}} V_i^{\min} \leq V_n,
    \end{aligned}
    \label{headroom}
\end{equation}
where $G_n$, $C_n$, $M_n$, and $V_n$ denote the total GPU accounts, CPU cores, RAM, and vRAM of server $n$, respectively.

Due to limited computational resources in practical edge networks, not all incoming prompts can be effectively served. A request may fail if the required LM is not currently deployed on any available node, or if resource contention leads to excessive queuing delays that exceed the service deadline. Therefore, the objective of this paper is to ensure fair and balanced service across multiple LM service types while reducing the end-to-end latency experienced by successfully processed requests. To achieve these objectives, we propose a joint optimization framework that integrates two performance criteria: (i) fairness across LM service types, to prevent persistent under-service of any type, and (ii) minimization of overall end-to-end latency for successfully processed requests.

Let $\rho_i$ denote the long-term service rate of service type $i \in \mathcal{I}$, defined as the fraction of type-$i$ requests that are successfully processed by the system. Specifically, the total number of arriving requests for service type $i$ over the entire horizon is
\begin{equation}
    A_i =
    \sum_{t \in \mathcal{T}} \sum_{n \in \mathcal{N}}
    \sum_{q \in \mathcal{Q}_{nt}} \mathbbm{1}\{i(q)=i\}
\end{equation}

To reflect the service-level behavior in real-world systems, we introduce an end-to-end deadline threshold $\tau$. A request is considered failed if it is not fully processed within this time limit, as delayed responses may violate quality-of-service (QoS) requirements. Accordingly, we define a binary indicator variable $\delta_q \in \{0,1\}$ to indicate whether request $q$ is successfully processed:
\begin{equation}
    \delta_q = 
\begin{cases}
1, & \text{if } i(q) \in \mathcal{I}^{\text{act}}_{\mathcal{D}_q} \text{ and } T_q \leq \tau, \\
0, & \text{otherwise},
\end{cases}
\end{equation}
where $\mathcal{D}_q$ is the server to which request $q$ is dispatched, $\mathcal{G}$ denotes the resource allocation policy, and \(T_q\) represents the end-to-end delay experienced by \(q\).

The number of successfully served type-$i$ requests is thus given by
\begin{equation}
    S_i =
    \sum_{t \in \mathcal{T}} \sum_{n \in \mathcal{N}}
    \sum_{q \in \mathcal{Q}_{nt}} \mathbbm{1}\{i(q)=i\}\delta_q,
\end{equation}
and the corresponding service rate is
\begin{equation}
    \rho_i = \frac{S_i}{A_i}, \quad \forall i \in \mathcal{I}.
\end{equation}

To quantify fairness across LM service types, we apply Jain’s fairness index~\cite{jain1999throughput} over the vector $\boldsymbol{\rho} = (\rho_1, \ldots, \rho_I)$:
\begin{equation}
    F(\boldsymbol{\rho}) = \frac{\left( \sum_{i \in \mathcal{I}} \rho_i \right)^2}{I \sum_{i \in \mathcal{I}} \rho_i^2},
\end{equation}
which lies in the interval $[{1}/{I}, 1]$, where a higher value indicates greater inter-type fairness. Here, fairness is evaluated across LM service categories. Each request in the service model is characterized by its required LM type, without additional user-identity or priority-class attributes.

For each successfully processed request $q$ (i.e., $\delta_q = 1$), the total end-to-end latency is composed of three components:
\begin{equation}
T_q =
L_q\left( G_{n_q, i(q)}, C_{n_q, i(q)} \right)
+ X_q(n_q)
+ Y_q(n_q, a_{n_q}),
\label{eq:latency_sum}
\end{equation}
where the first term $L_q(\cdot)$ captures the inference latency, which depends on the allocated GPU and CPU resources for LM type $i(q)$ on server $n_q$, as determined by the deployment policy $\mathcal{G}$. The second term $X_q(n_q)$ denotes the network transmission delay, including uplink and downlink delays between the origin node and target server $n_q$. The third term $Y_q(n_q, a_{n_q})$ represents the queuing delay, which depends on the current backlog and the LM deployment configuration at node $n_q$. $a_{n_q}$ denotes the realized LM deployment and resource allocation configuration at server $n_q$ when request $q$ enters service, as induced by the deployment policy $\mathcal{G}$.

To jointly optimize both service fairness and end-to-end latency, we normalize the two objectives onto a common scale. Fairness is normalized using a scaled variant of Jain's index:
\begin{equation}
    F_{\text{norm}}(\boldsymbol{\rho}) = \frac{F(\boldsymbol{\rho}) - \tfrac{1}{I}}{1 - \tfrac{1}{I}} \in [0, 1],
\end{equation}
so that $F_{\text{norm}} = 1$ corresponds to perfect fairness (i.e., all service types are equally served), and $F_{\text{norm}} = 0$ corresponds to the worst-case imbalance.

Latency is normalized based on the end-to-end deadline threshold $\tau$. 
\begin{equation}
\bar{T}_{\text{norm}} =
\frac{1}{\sum_{t \in \mathcal{T}} \sum_{n \in \mathcal{N}} \sum_{q \in \mathcal{Q}_{nt}} \delta_q}
\sum_{t \in \mathcal{T}} \sum_{n \in \mathcal{N}} \sum_{q \in \mathcal{Q}_{nt}}
\delta_q \cdot \frac{T_q}{\tau},
\end{equation}
This metric captures the average latency among completed requests. It isolates latency efficiency from the impact of failures, which are accounted for by the fairness objective.

Overall, the joint optimization problem is formulated as:
\begin{equation}
    P: \min_{\mathcal{D}, \mathcal{G}} \;
    \lambda \cdot \bar{T}_{\text{norm}} + (1 - \lambda) \cdot \left(1 - F_{\text{norm}}(\boldsymbol{\rho}) \right),
    \label{eq:obj}
\end{equation}
subject to server-level resource budgets Eq.~\eqref{headroom}, model deployment feasibility, and system-wide processing constraints. $\lambda \in [0,1]$ balances the trade-off between latency and fairness.

\vspace{-0.3cm}
\section{Methodology: Hierarchical Decision Framework for Joint Prompt Scheduling and Resource Allocation}
\label{sec:Methodology} 
\begin{figure*}[t]
    \centering
    \setlength{\abovecaptionskip}{0.2cm}
    \includegraphics[width=1\linewidth]{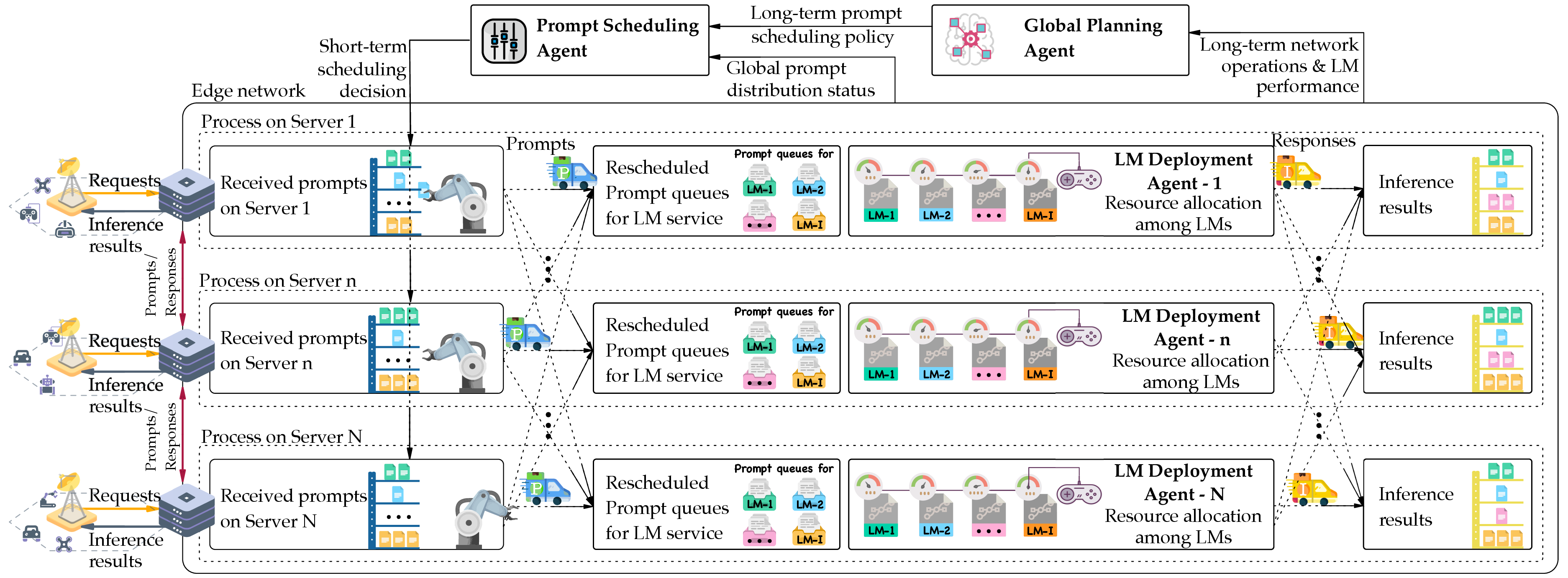}
    \caption{The architecture of the proposed Multi-Agentic AI solution for resource management of Edge LM inference.}
    \vspace{-0.5cm}
    \label{fig:algorithm}
\end{figure*}

To resolve this problem and enable efficient LM inference in edge networks, we propose a two-tiered Agentic AI-based optimization framework, as shown in Figure~\ref{fig:algorithm}. The upper tier features a long-horizon Global Planning Agent that determines traffic distribution and LM activation policies. The lower tier includes two components operating at each time slot: i) a Prompt Scheduling Agent for request dispatching and ii) multiple LM Deployment Agents deployed across the servers to manage model activation and resource allocation. 

The proposed hierarchy derives its decisions from runtime observations collected during system operation. Each edge-hosted LM is treated as an operational service described by recent latency, success ratio, queue evolution, and resource usage. The Global Planning Agent works with epoch-level summaries of these observations and previous policy outcomes, while the Prompt Scheduling Agent and LM Deployment Agents use current queue backlogs, deployment states, and resource headroom. This design allows the agents to coordinate heterogeneous LMs across changing model, hardware, and workload conditions. 
The hierarchy is further designed around the operating timescale of edge GenAI services. Pod and model initialization introduce second-level actuation overhead, making overly fine-grained reconfiguration inefficient, while queueing, model execution, and result delivery can span multiple slots. Request outcomes therefore become observable only after inference completes. The Global Planning Agent uses delayed performance feedback for long-horizon updates, whereas the Prompt Scheduling Agent and LM Deployment Agents react to current queue and resource states.

\vspace{-0.3cm}
\subsection{Tier-1: Long-term Global Planning Agent}
\label{method-tier1}
In real-world networks, the end-to-end latency $T_q$ and the outcome of a request $\delta_q$ are only observable after the entire inference process completes. 
As this process may span multiple time slots, it introduces substantial feedback delays, rendering traditional real-time control algorithms ineffective.
To address this challenge, the first tier of our solution introduces an LLM-based Global Planning Agent that operates over a long-term horizon (an epoch consisting of multiple time slots $\mathcal{T}_e$). Instead of relying on real-time feedback, it records the performance of completed requests over a retrospective window and stores this information in memory. By reasoning over both historical execution experience, the planner generates a macro-policy $\Pi_e$ aimed at minimizing the composite objective $\lambda \cdot \bar{T}_{\text{norm}} + (1 - \lambda) \cdot (1 - F_{\text{norm}})$. 

At the beginning of each epoch~$e$, the Planning Agent is informed by a comprehensive set of inputs, which are translated into a natural language prompt for the LLM. These inputs include:
\begin{itemize}[leftmargin=*]
    \item Objective description: The agent's objective is to generate a high-level policy that jointly governs prompt scheduling and LM deployment in an edge network, aiming to accelerate overall LM inference while ensuring fairness across service types, thereby avoiding the starvation of specific LMs.

    \item Static network context ($\mathbb{C}$): The heterogeneous resource capacities of all MEC servers, given by $\{G_n, C_n, M_n, V_n\}_{n \in \mathcal{N}}$, and the minimum memory requirements of all LM types, given by $\{M_i^{\min}, V_i^{\min}\}_{i \in \mathcal{I}}$, as defined in the problem formulation.

    \item Random policy baseline: A reference performance metric obtained from a uniformly randomized macro-policy. In cases where the current fairness score and mean end-to-end latency are lower than the established baseline, significant corrective actions are expected from the planner.

    \item LM-specific characteristics: A description of deployment constraints and capabilities specific to each LM type. For instance, certain models (e.g., for image generation) may be infeasible to run on CPU-based backends.

    \item Historical case memory ($\mathbb{H}$): A curated set of past experiences that form the in-context learning foundation for the LLM planner. Each case $h \in \mathbb{H}$ is structured as:
    \begin{equation}
        h = (\mathbb{M}_e, \Pi_e, F_{\text{norm}}(\boldsymbol{\rho}_e), \bar{T}_{\text{norm},e}),
        \label{eq:history}
    \end{equation}
    associating a past macro-policy $\Pi_e$ with its corresponding telemetry and performance.

    Here, the long-term telemetry $\mathbb{M}_e$ captures aggregated statistics over $\mathcal{T}_e$ for each service type $i \in \mathcal{I}$, including i) the success-only mean latency $\bar{T}_{i,e}$, computed over requests that were successfully processed within the queuing deadline and ii) the success ratio $\rho_{i,e}$, defined as the fraction of requests of type $i$ that were successfully completed. From these per-type metrics, we also derive the normalized global latency $\bar{T}_{\text{norm},e}$, representing the average latency across all successfully processed requests, scaled by the system deadline $\tau$ and the normalized Jain's fairness index $F_{\text{norm}}(\boldsymbol{\rho}_e)$, where $\boldsymbol{\rho}_e = (\rho_{1,e}, \ldots, \rho_{|\mathcal{I}|,e})$ denotes the vector of per-type success ratios during $\mathcal{T}_e$.
\end{itemize}

The output of the planning agent, the macro-policy $\Pi_e$, is a structured object comprising two key components: 
\begin{itemize}[leftmargin=*]
    \item Routing probabilities ($\mathcal{P}_e$): A set of probabilities
    \begin{equation}
        \mathcal{P}_e = \left\{ p_{in} \in [0,1] \;\middle|\; \sum_{n \in \mathcal{N}} p_{in} = 1,\ \forall i \in \mathcal{I} \right\},
    \end{equation} 
    which specifies the target traffic distribution for each service type $i$ across the MEC servers.

    \item Node role intent ($\mathcal{R}_e$): A set of role assignments
    \begin{equation}
    \mathcal{R}_e = \left\{ R_{n,e} \subseteq \mathcal{I} \;\middle|\; n \in \mathcal{N} \right\},
    \label{eq:nodes}
    \end{equation} 
    where each $R_{n,e}$ defines the subset of LM services that server $n$ is intended to host during epoch $e$.
\end{itemize}

Overall, drawing on principles from reinforcement learning, the approach avoids suboptimal macro-policy configurations by learning from negative precedents, while making targeted adjustments within optimal cases. This enables the Planning Agent to refine its decisions incrementally and achieve improved performance with respect to the composite objective.
In addition, empowered by in-context reasoning, it can generalize across varying traffic patterns and cluster conditions to produce robust global policies. 

\vspace{-0.3cm}
\subsection{Tier-2: Short-term Prompt Scheduling Agent and LM Deployment Agent}

While the  Global Planning Agent provides long-term strategic guidance by specifying target routing distributions and node roles, it operates at an epoch-level timescale and thus cannot fully capture the bursty and highly dynamic nature of real-world edge workloads. In practice, queue lengths, instantaneous load, and resource availability can fluctuate significantly from one time slot to the next. To bridge this gap between slow-timescale planning and fast-timescale dynamics, Tier-2 introduces two complementary agents: a Prompt Scheduling Agent that performs slot-level prompt dispatching under the planner's guidance, and multiple distributed LM Deployment Agents that continuously adapt local LM activation and resource allocation on each MEC server.

\subsubsection{Prompt Scheduling Agent}
To handle the burstiness and unpredictability of network traffic, we design an LLM-based \textit{Prompt Scheduling Agent} that operates at every time slot $t$, translating the high-level strategy of the Global Planning Agent into concrete routing decisions based on current network and edge-server conditions. 
Specifically, the scheduler observes: (i) the macro-level strategy $\Pi_e$, which specifies per-LM routing probabilities and node-role assignments (i.e., which LM types are allowed to run on which nodes); and (ii) a short-term system snapshot $\mathbb{S}_t$ that captures dynamic conditions such as per-node prompt backlogs $\{Q_{ni}(t)\}_{i \in \mathcal{I},\, n \in \mathcal{N}}$, where $Q_{ni}(t)$ denotes the total number of prompts of type $i$ currently waiting at server $n$.
For each incoming request $q \in \mathcal{Q}_{nt}$, the scheduler determines a dispatch decision $\mathcal{D}_q \in \mathcal{N}$, thereby filling a routing matrix that maps each $(n, i(q))$ pair, i.e., origin node $n$ and request type $i(q)$, to a target MEC server for inference execution.

While the scheduler primarily follows the probabilistic guidance specified in $\mathcal{P}_e$, maintaining each LM's long-term traffic split close to the macro-level distribution, it also dynamically adjusts routing decisions based on the instantaneous cluster state $\mathbb{S}_t$.
For example, if the preferred target node $n^*$ for service $i(q)$ is currently overloaded (e.g., exhibiting deep queues or low available GPU headroom), the agent may temporarily reroute the request to an alternative feasible node, as long as the decision remains consistent with the node-role intent defined in $\mathcal{R}_e$. This mechanism enables timely load balancing and congestion avoidance while respecting the global policy structure.

\subsubsection{LM Deployment Agent}
To further optimize the system-wide objective, reducing end-to-end latency and improving fairness across LM service types, we introduce distributed \textit{LM Deployment Agents}, invoked locally on each server $n \in \mathcal{N}$. Their objective is to reduce inference and queuing delay, while satisfying server-level resource budgets and deployment feasibility conditions defined in the problem formulation.

To support adaptive resource orchestration under the dynamically shifting prompt distribution policies of the Planning Agent, the LM Deployment Agents are also powered by LLMs, which are capable of generalizing across diverse and non-stationary request patterns induced by evolving global macro-policies $\Pi_e$.
Unlike traditional AI-based solutions that may overfit to specific traffic scenarios, LLM can holistically interpret high-level deployment intent and local runtime states to make informed and flexible decisions.

Each on-node agent observes local MEC information, including per-LM queue backlogs, current LM deployment status $z_{ni}(t)$, and resource headroom in terms of CPU, memory, and available GPU replicas. Guided by the macro-level node-role intent (Eq.~\eqref{eq:nodes}), these agents operate on the same short-term timescale as the Prompt Scheduling Agent and make fine-grained resource management decisions, including selecting the active LM instances $\mathcal{I}_n^{\text{act}}$ and determining the corresponding CPU and GPU allocations $a_n(t)$, according to instantaneous workload fluctuations and queue dynamics.

The Tier-2 agents continuously execute per-request routing and per-node deployment actions, which collectively determine end-to-end latency and service success outcomes for each LM type. Over fixed evaluation windows, these outcomes are aggregated into per-type and global performance metrics, such as success-only mean latency and service rates, which are fed back to the Tier-1 Planning Agent as structured telemetry. This closed-loop design enables Tier-1 to periodically update the macro-policy $\Pi_e$ based on empirical evidence from Tier-2 behaviors, while Tier-2 adapts its decisions immediately in response to the latest strategy.

\subsubsection{Stability Conditions and Two-Timescale Interpretation} 
\label{sec:stability}

The proposed framework separates epoch-level global planning from slot-level local control. The Tier-1 Planning Agent updates the macro-policy $\Pi_e$ once per epoch, while Tier-2 agents execute prompt routing and LM deployment decisions at the slot level. This separation allows the lower layer to react to short-term queue variations within an epoch before the upper layer updates its long-horizon policy.

Consider an epoch $e$ in which $\Pi_e$ is held constant. Define the quadratic backlog potential
\begin{equation}
    \Phi(t)=\frac{1}{2}\sum_{n\in\mathcal{N}}\sum_{i\in\mathcal{I}}Q_{ni}^{2}(t),
\end{equation}
where $Q_{ni}(t)$ denotes the backlog of type-$i$ prompts at server $n$. Suppose the arrival-rate vector induced by $\Pi_e$ lies inside the stabilizable service region, and per-slot arrivals and departures are bounded. Suppose further that the Tier-2 routing and deployment decisions provide sufficient queue-draining service so that the conditional drift satisfies
\begin{equation}
    \mathbb{E}\!\left[\Phi(t+1)-\Phi(t)\mid \mathbf{Q}(t)\right]
    \leq B-\epsilon\sum_{n,i}Q_{ni}(t),
\label{eq:drift_sufficient}
\end{equation}
where $\mathbf{Q}(t)=\{Q_{ni}(t)\}_{n\in\mathcal{N},i\in\mathcal{I}}$ is the full backlog vector, $B$ is a finite drift constant resulting from bounded per-slot arrivals and departures, and $\epsilon$ is the queue-draining service margin. Under this condition, standard Lyapunov-drift arguments imply that the long-run time-averaged aggregate backlog is bounded, with an upper bound proportional to $B/\epsilon$~\cite{neely2010stochastic}.

In the proposed framework, the Prompt Scheduling Agent observes queue pressure when selecting destinations, while each LM Deployment Agent observes local backlog, activation state, and resource headroom. These observations allow Tier-2 to respond to queue pressure, while deterministic validation removes infeasible actions. However, neither mechanism guarantees that the drift condition is satisfied. Therefore, we present Equation~\eqref{eq:drift_sufficient} only as a conditional stability interpretation. If the realized routing and deployment actions induce the stated negative drift, the long-run time-averaged backlog is bounded. Under this stability condition, and provided that the Tier-2 dynamics evolve on a sufficiently faster timescale than the Tier-1 updates, epoch-level aggregation can attenuate transient slot-level fluctuations~\cite{borkar2008stochastic}. This aggregated telemetry provides a more comparable summary of the latency and fairness induced by each macro-policy and, together with episodic memory, informs subsequent Tier-1 policy refinement. The closed-loop behavior is empirically characterized by the objective trajectories in Figure~\ref{fig:learning-dynamics}. 

\vspace{-0.3cm}
\subsection{Observation and Action Spaces}
\label{subsec:obs-action-spaces}

The preceding descriptions can be summarized as the following observation--action interfaces. At epoch~$e$, the Global Planning Agent observes
\begin{equation}
o_e^{\mathrm{G}} =
\bigl(
\mathbb{C},\,
\mathbb{H}_e,\,
\mathbb{M}_{e-1},\,
\Pi_{e-1}
\bigr).
\label{eq:global-observation}
\end{equation}
where $\mathbb{C}$ is the static cluster and LM context, $\mathbb{H}_e$ is the retrieved historical case memory, $\mathbb{M}_{e-1}$ contains the latest epoch-level telemetry, and $\Pi_{e-1}$ is the previously applied macro-policy. Its action is the next macro-policy
\begin{equation}
\begin{aligned}
a_e^{\mathrm{G}}
&= \Pi_e
= \bigl(\mathcal{P}_e,\mathcal{R}_e\bigr), \\
\mathcal{P}_e
&=
\{p_{in,e}\}_{i\in\mathcal{I},\,n\in\mathcal{N}}, \\
\mathcal{R}_e
&=
\{R_{n,e}\subseteq\mathcal{I}\}_{n\in\mathcal{N}} .
\end{aligned}
\label{eq:global-action}
\end{equation}
where $\sum_{n\in\mathcal{N}}p_{in,e}=1$ for each LM type $i$. At slot~$t$ within epoch~$e$, the Prompt Scheduling Agent observes
\begin{equation}
\begin{aligned}
o_t^{\mathrm{S}}
={}&
\bigl(
\Pi_e,\,
\mathbb{S}_t,\,
\mathcal{Q}_t
\bigr), \\
\mathbb{S}_t
={}&
\bigl(
	\{Q_{ni}(t)\},\,
\{z_{ni}(t)\}, \\
&\quad
\{\eta_n^{\mathrm{cpu}}(t),
\eta_n^{\mathrm{mem}}(t),
\eta_n^{\mathrm{gpu}}(t)\}
\bigr).
\end{aligned}
\label{eq:scheduling-observation}
\end{equation}
where $Q_{ni}(t)$ is the type-$i$ backlog at server $n$, $z_{ni}(t)$ denotes the current activation state of LM~$i$ at server~$n$, $\eta_n^{\cdot}(t)$ denotes residual resource headroom, and $\mathcal{Q}_t=\bigcup_{n\in\mathcal{N}}\mathcal{Q}_{nt}$. Its action is a slot-level routing map that defines per-request dispatch decisions
\begin{equation}
a_t^{\mathrm{S}}
=
\{d_{ni}(t)\in\mathcal{N}\}_{n\in\mathcal{N},\,i\in\mathcal{I}},
\
\mathcal{D}_q(t)=d_{n,i(q)}(t),\; q\in\mathcal{Q}_{nt}.
\label{eq:scheduling-action}
\end{equation}
where each per-request dispatch decision $\mathcal{D}_q(t)$ must satisfy the LM placement and hardware feasibility constraints described above. Finally, each LM Deployment Agent at server~$n$ observes
\begin{equation}
\begin{aligned}
o_{n,t}^{\mathrm{D}}
={}&
\bigl(
R_{n,e},\,
	\{Q_{ni}(t)\}_{i\in\mathcal{I}},\,
\{z_{ni}(t)\}_{i\in\mathcal{I}}, \\
&\quad
\eta_n^{\mathrm{cpu}}(t),\,
\eta_n^{\mathrm{mem}}(t),\,
\eta_n^{\mathrm{gpu}}(t)
\bigr).
\end{aligned}
\label{eq:deployment-observation}
\end{equation}
and selects a feasible local deployment action
\begin{equation}
\begin{aligned}
a_{n,t}^{\mathrm{D}}
&= a_n(t) \\
&=
\bigl(
\mathcal{I}_n^{\mathrm{act}}(t),\,
\{C_{ni}(t),G_{ni}(t)\}_{i\in\mathcal{I}_n^{\mathrm{act}}(t)}
\bigr)
\in \mathcal{A}_{n}^{\mathrm{D}}.
\end{aligned}
\label{eq:deployment-action}
\end{equation}
where $a_n(t)$ specifies the active LM set and the CPU/GPU allocation for each active LM, and $\mathcal{A}_{n}^{\mathrm{D}}$ contains only local configurations satisfying the resource constraints in Eq.~\eqref{headroom} and transient deployment-state feasibility. The natural-language prompts used by the agents are deterministic serializations of these tuples, while the returned control objects or placement statements are mapped to the corresponding action spaces before execution.
Based on these interfaces, Algorithm~\ref{alg:ma-runtime} summarizes the coordination among epoch-level planning, slot-level prompt scheduling, and on-node LM deployment in the proposed framework.

\begin{algorithm}[t]
\caption{Decision and validation logic of the proposed Multi-Agentic AI framework.}
\label{alg:ma-runtime}
\begin{algorithmic}[1]
\State \textbf{Input:} Cluster context $\mathbb{C}$, historical memory $\mathbb{H}$, epoch length $|\mathcal{T}_e|$, feasible deployment sets $\{\mathcal{A}_n^{\mathrm{D}}\}_{n\in\mathcal{N}}$.
\State \textbf{Initialize:} Macro-policy $\Pi_0$, monitoring database, and per-node LM deployment states.
\For{each epoch $e=1,2,\ldots$}
    \State Collect completed-request telemetry $\mathbb{M}_{e-1}$ and retrieve relevant memory cases $\mathbb{H}_e$.
    \State Serialize $o_e^{\mathrm{G}}=(\mathbb{C},\mathbb{H}_e,\mathbb{M}_{e-1},\Pi_{e-1})$ into the Planning Agent prompt.
    \State Call the Planning Agent to produce $\Pi_e=(\mathcal{P}_e,\mathcal{R}_e)$.
    \State Validate $\Pi_e$ for probability normalization, LM placement feasibility, and node-capacity consistency; if validation fails, use a safe fallback policy.
    \For{each slot $t\in\mathcal{T}_e$}
        \State Collect $\mathbb{S}_t$ from monitoring data available at the end of slot $t-1$.
        \State Serialize $o_t^{\mathrm{S}}=(\Pi_e,\mathbb{S}_t,\mathcal{Q}_t)$ and call the Prompt Scheduling Agent.
        \State Validate the slot-level routing map $\{d_{ni}(t)\}$ against LM placement and hardware feasibility constraints, then dispatch requests according to $\mathcal{D}_q(t)=d_{n,i(q)}(t)$.
        \For{each server $n\in\mathcal{N}$ in parallel}
            \State Build $o_{n,t}^{\mathrm{D}}$ from the node role, local backlogs, activation states, and resource headroom.
            \State Call the local LM Deployment Agent to produce $a_n(t)\in\mathcal{A}_n^{\mathrm{D}}$.
            \State Validate $a_n(t)$ against resource budgets and deployment-state constraints, then apply the corresponding deployment operations.
        \EndFor
        \State Execute inference, record latency components and success indicators, and append them to telemetry memory.
    \EndFor
\EndFor
\end{algorithmic}
\end{algorithm}

The deterministic control-plane complexity is modest relative to LM inference and model deployment time. At the epoch level, validating the macro-policy scans the $|\mathcal{I}|\times|\mathcal{N}|$ routing-probability table and checks the node-role intents over LM--node pairs, giving $O(|\mathcal{I}||\mathcal{N}|)$ deterministic work per epoch, excluding the external LLM API call. At each slot, constructing $\mathbb{S}_t$ scans per-LM queue and activation states across nodes and resource headroom, validating the routing map checks one destination for each origin-node/LM-type entry, and applying the map assigns destinations to the arriving requests. This gives $O(|\mathcal{I}||\mathcal{N}|+|\mathcal{Q}_t|)$ deterministic work per slot, where $|\mathcal{Q}_t|$ is the number of arriving requests in slot $t$. For LM deployment, each server validates the returned per-LM placement vector by scanning LM modes and resource totals, costing $O(|\mathcal{I}|)$ per server. Thus, the sequential aggregate slot-level deterministic cost is $O(|\mathcal{I}||\mathcal{N}|+|\mathcal{Q}_t|)$, while the local deployment checks run in parallel across servers.
These interfaces also define the boundary of LLM-generated decisions. The agents return structured control objects corresponding to the macro-policy, routing, and local deployment action spaces defined above. LLM reasoning is therefore confined to policy, routing, and placement choices, while state collection and control execution are handled by deterministic system logic.

Overall, our solution provides a modular, adaptive, and fully deployable framework for optimizing LM inference in mobile edge networks. By leveraging agentic AI modules for both long-term planning and short-term control, the system achieves scalable coordination across servers and responsive adaptation within individual MEC nodes under dynamic traffic loads and heterogeneous service demands. Moreover, all agents in our solution are dedicated solely to infrastructure orchestration and operate exclusively on system-level telemetry. In contrast to the user-facing LMs that process multimodal and private input content, these agents do not access or process end-user data. This separation preserves privacy and enables practical deployment via cloud-hosted LLM APIs, without violating data governance requirements. In this way, our framework exemplifies the synergy of \textit{GenAI for Networks}, using generative agents to manage distributed infrastructure, and \textit{Networks for GenAI}, adapting edge systems to better serve edge GenAI workloads.

\vspace{-0.3cm}
\section{Experiment Setup and Implementation}
\label{sec:testbed}
To evaluate the effectiveness of the proposed solution, we deploy a network with multiple servers distributed across Bristol. The setup of the network and the deployment of the two-tiered solution are detailed in this section{\footnote{Multi-agentic AI solution, multi-model LM Dockerfiles, and network configurations are available at \url{https://github.com/hpn-bristol/agentic-edge-lm}. Deploying the framework on a new testbed requires following the network setup and containerized LM deployment procedures described in this section.}}.

\vspace{-0.4cm}
\subsection{Network architecture}

\begin{figure}[t]
    \centering    
    \setlength{\abovecaptionskip}{0.2cm}
    \includegraphics[width=1\linewidth]{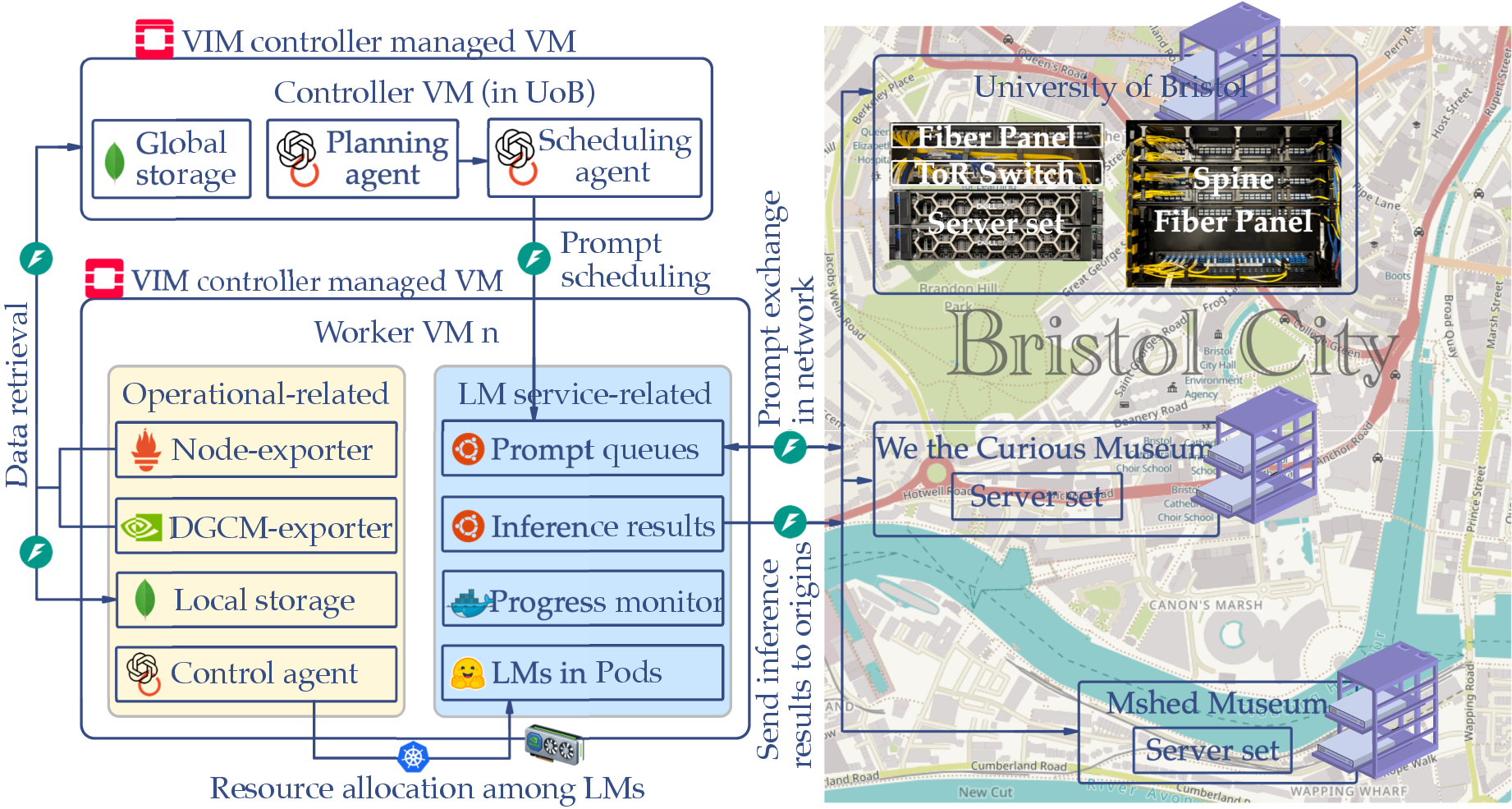}
    \caption{Architecture of the testbed and the allocation of network elements among servers.}
    \vspace{-0.3cm}
    \label{fig:testbed}
\end{figure}
Figure~\ref{fig:testbed} illustrates the network architecture, with three server sets distributed across the University of Bristol, the We the Curious Museum, and the Mshed Museum. 
In this network, servers within a rack are connected to the Top-of-Rack (ToR) switch via 100 Gigabit Ethernet links, which are then linked to a fiber patch panel using 100Gbps Small Form-factor Pluggable (SFP) transceivers. At the intra-location level, the fiber patch panel connects to an aggregation switch, aggregating and managing traffic from all racks within the same location. For inter-location connectivity, the aggregation switch links to a spine fiber patch panel, enabling communication between aggregation switches across different locations. A router is implemented at the University of Bristol site to enable connectivity between the internal network and the external internet.

Seven Virtual Machines (VMs) are deployed based on OpenStack~\cite{sefraoui2012openstack} across these locations, each serving as a MEC node. These VMs are interconnected through a network-layer mesh in which every VM can establish Secure Shell (SSH) connectivity with the remaining nodes.  A Kubernetes cluster is implemented within this infrastructure to facilitate efficient management of the distributed system and the execution of LM services. Table~\ref{configuration} summarizes the configurations of these VMs. 
Among them, VM1 is designated as the control node and is used exclusively for the centralized Global Planning Agent and Prompt Scheduling Agent. The remaining worker nodes (VM2--VM7) execute the LM Deployment Agents and LM inference workloads. 

\begin{table}[t]
  \centering
  \caption{Configurations of VMs in the experiment network}
  \scalebox{0.91}{
    \begin{tabular}{l@{\hspace{0.5cm}}c@{\hspace{0.5cm}}c@{\hspace{0.5cm}}c@{\hspace{0.5cm}}c}
    \toprule
          & CPU (model-cores)   & RAM   & GPU   & vRAM \\
    \midrule
    VM 1  & Intel Xeon Icelake - 8  & 16GB   & N/A & N/A \\
    VM 2  & AMD EPYC 7453 - 24 & 32GB   & N/A & N/A \\
    VM 3  & AMD EPYC 7453 - 16  & 24GB   & NVIDIA A5000 * 1  & 24GB \\
    VM 4  & AMD EPYC 7453 - 16  & 24GB   & NVIDIA A5000 * 1  & 24GB \\
    VM 5  & AMD EPYC 7453 - 16  & 24GB   & NVIDIA A5000 * 1  & 24GB \\
    VM 6  & AMD EPYC 7453 - 16  & 24GB   & NVIDIA A5000 * 1  & 24GB \\
    VM 7  & AMD EPYC 7453 - 16  & 24GB   & NVIDIA A5000 * 2  & 48GB \\ 
    \bottomrule
    \end{tabular}%
    }
\label{configuration}%
\end{table}%

\vspace{-0.4cm}
\subsection{Containerized LM deployment}
Within each VM, LMs can be deployed in dedicated Kubernetes pods, with one LM per pod. To minimize the initialization latency of LMs and avoid redundant LM parameter downloads when a pod restarts, we utilize Dockerfiles to encapsulate each LM into an image. This image includes the base execution environment (e.g., operating system), application code for each LM, LM execution dependencies (e.g., libraries), and pre-cached files (e.g., LM parameters). The resulting images are pushed to a local Docker registry for reuse during future pod creations. Consequently, when a new pod is created, the underlying container can be directly deployed from the pre-built image. While this approach significantly reduces the deployment time of LM from scratch, the initialization of pods and the transfer of LM parameters from storage to memory or vRAM cannot be achieved in real time. These initialization times are incorporated into the optimization by considering them in the queuing delay $Y_q$.

Four multi-modal LMs (i.e. $I = 4$) are selected to assess the performance of the proposed solution, covering four types of common requirements, including i) LM1: simple text-to-text using GPT2 with 137M parameters~\cite{radford2019language}, ii) LM2: complex text-to-text using GPT2-large with 812M parameters~\cite{radford2019language}, iii) LM3: image-to-text using BLIP with 470M parameters~\cite{Junnan2024blip} and iv) LM4: text-to-image using Stable Diffusion with 890M parameters~\cite{Rombach_2022_CVPR}. \textit{low\_cpu\_mem\_usage} from \textit{accelerate} developed by Huggingface is applied to minimize memory usage. Based on empirical testing, the minimum memory/storage requirements for the containers hosting these four LMs are 1.2Gb/7.5Gb, 6.5Gb/13Gb, 0.5Gb/10Gb, and 1.6Gb/17.5Gb, respectively. To accommodate LM caching and to prevent memory exhaustion in pods, the deployment memory for these LMs has been configured at 3GB, 9GB, 2.5GB, and 3.5GB. 
At each time slot $t$, each MEC server receives up to four LM service requests, i.e., $|\mathcal{Q}_{nt}| \leq 4$. The number of prompts per request is set as $K_q \in \{0, 1, \ldots, 8\}$.

\vspace{-0.3cm}
\subsection{Model parallelism on GPU}
The \textit{NVIDIA Container Toolkit} is set up on all VMs, which allows users to build and run GPU-accelerated containers by providing a container runtime library and utilities that automatically configure containers to leverage NVIDIA GPUs.  Subsequently, the \textit{NVIDIA device plugin} is deployed as a \textit{DaemonSet} on all VMs, allowing Kubernetes to discover and manage GPU resources.
To enable parallel computation of multiple LM across various containers, Nvidia provides two options: Multi-Instance GPU (MIG) and time-slicing. MIG enables static partitioning of GPUs into multiple isolated instances, each with its own dedicated portion of resources, such as vRAM and streaming multiprocessors (SMs). In contrast, time slicing divides the GPU’s time into intervals and assigns these time slots to various workloads or VMs based on predefined policies. During each time slice, a specific workload has exclusive access to GPU resources, while vRAM remains isolated among processes.
Since MIG is only applicable to certain GPU models, the time-slicing strategy is adopted to ensure algorithm generality. It is implemented by using \textit{ConfigMap} and each physical GPU is split into 2 vGPUs. When the LM is executed on multiple vGPUs, we employ an automatic \textit{device\_map} mechanism based on \textit{transformer} developed by Huggingface, which automatically distributes the model’s weights across the allocated vGPUs, enabling parallel computation and maximizing resource efficiency.

\vspace{-0.4cm}
\subsection{Network monitoring \& management }
The interaction between the two-tiered solution and the network involves two main components: i) network monitoring, which encompasses the observation of network status and processing delays; and ii) network management, which pertains to the realization of Agentic AI and the management actions within the network.

\subsubsection{Network monitoring}
A \textit{Prometheus Node Exporter} and a \textit{nvidia-dcgm-exporter} instance are deployed on each VM within the cluster using a \textit{Daemonset} to monitor CPU, RAM, GPU and vRAM metrics of each node. 
To obtain real-time network information from \textit{Prometheus Node Exporter} and \textit{nvidia-dcgm-exporter}, the controller VM dispatches concurrent HTTP requests to the \textit{/metrics} RESTful API endpoints of all servers, utilizing ports 9100 and 9400 as the defaults for CPU and GPU-related metrics, respectively. This approach enables the retrieval of monitored network information across the cluster. The collected data is subsequently stored in a global MongoDB database for the inference purposes of the Global Planning Agent and the Prompt Scheduling Agent. On each worker node, a similar strategy is employed, where each server saves its own resource information in a local MongoDB database. This local storage gives each LM Deployment Agent rapid access to node-specific metrics without introducing unnecessary network transfers.  As a prerequisite, \textit{Role-Based Access Control (RBAC)} configurations are required to ensure that the pods have the necessary permissions to query and collect metrics and access the MongoDB endpoint. 
To calculate the delayed feedback $T_q$ and $\delta_q$, the uplink time, queuing time, inference time, and downlink time of each prompt set are recorded in the global MongoDB database. Each entry is annotated with the corresponding step and additional relevant metadata to enable efficient retrieval by the agents.

\subsubsection{Network management}
The proposed Multi-Agentic AI framework comprises a total of $8$ agents, all built upon the OpenAI API. Specifically, a Global Planning Agent and a Prompt Scheduling Agent are invoked from VM1, while six LM Deployment Agents are each called via API from their respective MEC nodes during execution.

During runtime, these agents operate periodically within a shared slot-level control cycle. Based on pre-benchmark measurements of deployment actuation, request completion, and agent response latency, we use a nominal $30$-second slot. This choice reflects three timing requirements. First, the slot must be long enough to accommodate deployment actuation and Kubernetes state propagation, avoiding another deployment decision while the previous activation or deployment transition is still taking effect. In $80$ pre-benchmark cold-start and deployment-transition measurements, the delay from request generation to inference start had a median of $9.34$\,s, a $95$th percentile of $12.23$\,s, and a maximum of $13.58$\,s. Second, the slot must remain short enough to incorporate newly completed request outcomes into subsequent routing and deployment decisions. Of the $344$ successful requests in the pre-benchmark service profile, the median end-to-end completion time was $9.76$\,s, and $288$ requests ($83.72\%$) completed within $30$\,s. A longer interval would defer the use of this feedback and delay the corresponding control updates. Third, both the LM Deployment Agent control process across all workers and the Prompt Scheduling Agent call should normally complete within the slot. Across more than $3{,}000$ LM Deployment Agent control cycles, the elapsed time from request generation to the final deployment-state snapshot across all workers had a median of $3.71$\,s and a $95$th percentile of $7.93$\,s, with $99.82\%$ of the cycles completing within $30$\,s. Separately, the Prompt Scheduling Agent call had median and $95$th-percentile response latencies of $2.74$\,s and $4.50$\,s, respectively. The slot is a nominal control interval instead of a hard timeout, allowing a rare overrun to complete before the next slot begins. 

At Tier‑1, the Long-term Global Planning Agent runs once per epoch ($50$ slots). We set $\tau=900\,\mathrm{s}$ as a request timeout to prevent long-running or stalled requests from occupying worker resources indefinitely. The threshold is applied to all LM types and methods and does not represent differences in the latency requirements of the four LM types. To evaluate this setting, we conducted preliminary trials with random prompt routing and LM deployments. Across $9{,}453$ requests covering all four LM types, $6{,}365$ requests ($67.3\%$) completed within $900$\,s. Therefore, the timeout covers the majority of requests even under uncoordinated control while limiting the prolonged tail caused by deployment mismatches and queue accumulation. Moreover, $900$\,s corresponds to a waiting time of $15$ minutes and may already exceed an acceptable user experience for interactive LM inference.  The weighting parameter is set to $\lambda = 0.5$, so that both end-to-end latency and fairness across LM service types are treated with equal importance in the optimization objective.

 The numerical statistics in Eq.~\eqref{eq:history} are rendered into natural‑language sentences, for example: \textit{“In the last epoch, LM1/LM2/LM3/LM4 have success‑only mean latencies $210.5, 245.3, 120.7, 680.2,$s with success ratios $0.93, 0.88, 0.97, 0.72$, respectively. The resulting global mean latency is $314.2,$s and the normalized Jain fairness over success ratios is $0.81$. The corresponding macro policy allocated LM1/LM2 mostly to worker1–2, LM3/LM4 to worker3–5, and the micro router had an off‑role ratio of $9.5\%$.”} This textual summary is concatenated with a role description that states the objective (\textit{“reduce global average latency while improving fairness across LMs’ success service ratios”}) and the static cluster description. 

 The Planning Agent is instructed to output a strict JSON object containing: (i) a routing probabilities section that, for each LM, specifies a soft distribution of traffic across MEC nodes and (ii) a node role intent section describing, for each node, which LM types it is allowed or expected to host. The JSON is validated to ensure non‑negative weights that sum to one over the feasible node set for each LM, that GPU‑only models are only assigned to GPU‑capable nodes, and that per‑node LM sets respect concurrent capacity limits; if parsing fails or constraints are violated, we fall back to a random baseline policy. The validated macro policy is persisted in a shared store for consumption by the two lower‑layer agents.
Given that the Planning Agent learns from historical execution traces to generate refined decisions and operates without strict latency requirements, we employ a GPT-5-based API to leverage its advanced reasoning capacity.

At Tier-2, the Prompt Scheduling Agent is implemented as an asynchronous Python service that runs one control step per slot. At the beginning of each slot $t$, immediately after slot $t-1$ has finished, it loads the latest macro policy JSON file, queries the monitoring layer for a short‑window snapshot of cluster state (per‑node queue backlogs, CPU/memory/GPU headroom, and current LM activations) aggregated up to the end of slot $t-1$, and summarizes this information into a compact natural‑language prompt. This prompt describes both the Planning Agent’s intended long‑term traffic splits and the current short‑term load imbalances and is sent to an OpenAI API.

It returns a JSON-formatted routing plan that maps each origin-node and LM-type pair to a preferred destination node.  The plan is then used for all requests arriving during slot $t$ and broadcast to worker nodes via a FastAPI‑based control plane. The routing output is accepted only in the predefined JSON schema. Malformed or infeasible routing outputs are not used and are replaced by local routing for that slot. On each node, the prompt‑scheduling service is encapsulated within a \textit{DaemonSet} pod. A \textit{port‑forward} Kubernetes service is configured between the host VM and the internal Kubernetes network to deliver the JSON messages into the appropriate \textit{DaemonSet} pods, where they are translated into per‑request forwarding actions.

In addition, the LM Deployment Agent on each server is implemented with a short system prompt specialized to that node’s hardware profile (CPU‑only, single‑GPU, or dual‑GPU). Rather than inputting the full macro policy, these agents receive a compact node‑role description derived from the Planning Agent’s node role intent, indicating which LM types are primarily expected to run on that node. At the beginning of each local control step, synchronized with the global slot boundary and using measurements accumulated up to the end of the previous slot, each agent inspects a structured summary of its own state, including per‑LM queue backlogs on that node, current pod/VM deployment status, and approximate CPU/memory/GPU headroom, and then selects a discrete activation pattern over LM types (e.g., for each LM: \emph{on GPU}, \emph{on CPU}, or \emph{off}), subject to local feasibility constraints. The local deployment output is restricted to a fixed-format placement statement and mapped to a feasible activation pattern in $\mathcal{A}_n^{\mathrm{D}}$ before execution, while API failures or uninterpretable responses fall back to a feasible placement from the same local action space.

 The resulting activation pattern is encoded into concrete start/stop operations for LM replicas on that node (e.g., creating or terminating Kubernetes pods with appropriate CPU/GPU bindings). 
Because each LM Deployment Agent is co-located with the edge LM inference application, the deployment decision is made locally without inter-node transmission.  In the creation and updating operations, CPU resources for each pod are managed by specifying \textit{CPU requests} and \textit{CPU limits} in the Kubernetes deployment configuration, both set to the same value as determined by the LM Deployment Agent. \textit{CPU requests} represent the minimum guaranteed CPU resources for a pod and determine its CPU shares when resources are contested, while \textit{CPU limits} set the maximum CPU resources a pod can use. These Kubernetes settings are directly translated into Control Groups (cgroups) configurations on the node where the pod runs. CPU requests are converted into the \textit{cpu.shares} parameter within cgroups, which determines the pod's relative share of CPU time under resource contention.  The \textit{CPU limits} are enforced through the \textit{cpu.cfs\_quota\_us} and \textit{cpu.cfs\_period\_us} parameters in cgroups, defining the maximum CPU time a pod can consume within a specific period. The \textit{Linux scheduler} allocates CPU time to processes based on these cgroup settings. By adhering to the defined shares (\textit{cpu.shares}) and quotas (\textit{cpu.cfs\_quota\_us}), the \textit{Linux scheduler} ensures a deterministic distribution of CPU resources according to the specified agent actions.

 As for GPU resources, they are managed by \textit{Kubernetes Device Plugin} by specifying \textit{nvidia.com/gpu} indicator in resource requests and limits configurations. Different from CPUs, which can be fractionally allocated, GPUs are indivisible. They can only be virtualized into a discrete number of vGPUs. The LM Deployment Agent determines the number of vGPUs allocated to each LM. 
In Tier-2, since all agents only need to follow the high-level policy provided by the Planning Agent and apply light-touch adjustments based on the current network state, their tasks are relatively simple. Moreover, as these decisions operate at a short-term timescale, we adopt a GPT-4-based API for their implementation.

It is important to note that deactivating an LM pod may interrupt in-progress requests in the local queue, resulting in partial or failed processing. Moreover, due to the substantial size of LMs, the termination process is not instantaneous. During this shutdown period, the occupied resources (e.g., CPU, memory, or GPU) are not yet released, potentially causing resource contention and delaying subsequent deployments by placing them into a Pending state. 
To mitigate these disruptions, we implement two strategies: 
\begin{itemize}[leftmargin=*]
    \item To enable breakpoint recovery following Pod reactivation and to prevent redundant processing of prompts within the same execution group, we employ a \textit{hostPath} volume to store a CSV file on the VM. This file maintains a flag indicating the real-time processing state, enabling the internal Pod to resume execution from its last recorded position and ensuring computational continuity.
    \item To prevent conflicts and resource exhaustion, we enforce a strict rule: if a new Pod is created but remains pending, any subsequent action requiring additional resource allocation is automatically voided.
\end{itemize}

\vspace{-0.4cm}
\subsection{Runtime operation}

\begin{figure}[t]
    \centering
    \setlength{\abovecaptionskip}{0.2cm}
    \includegraphics[width=1\linewidth]{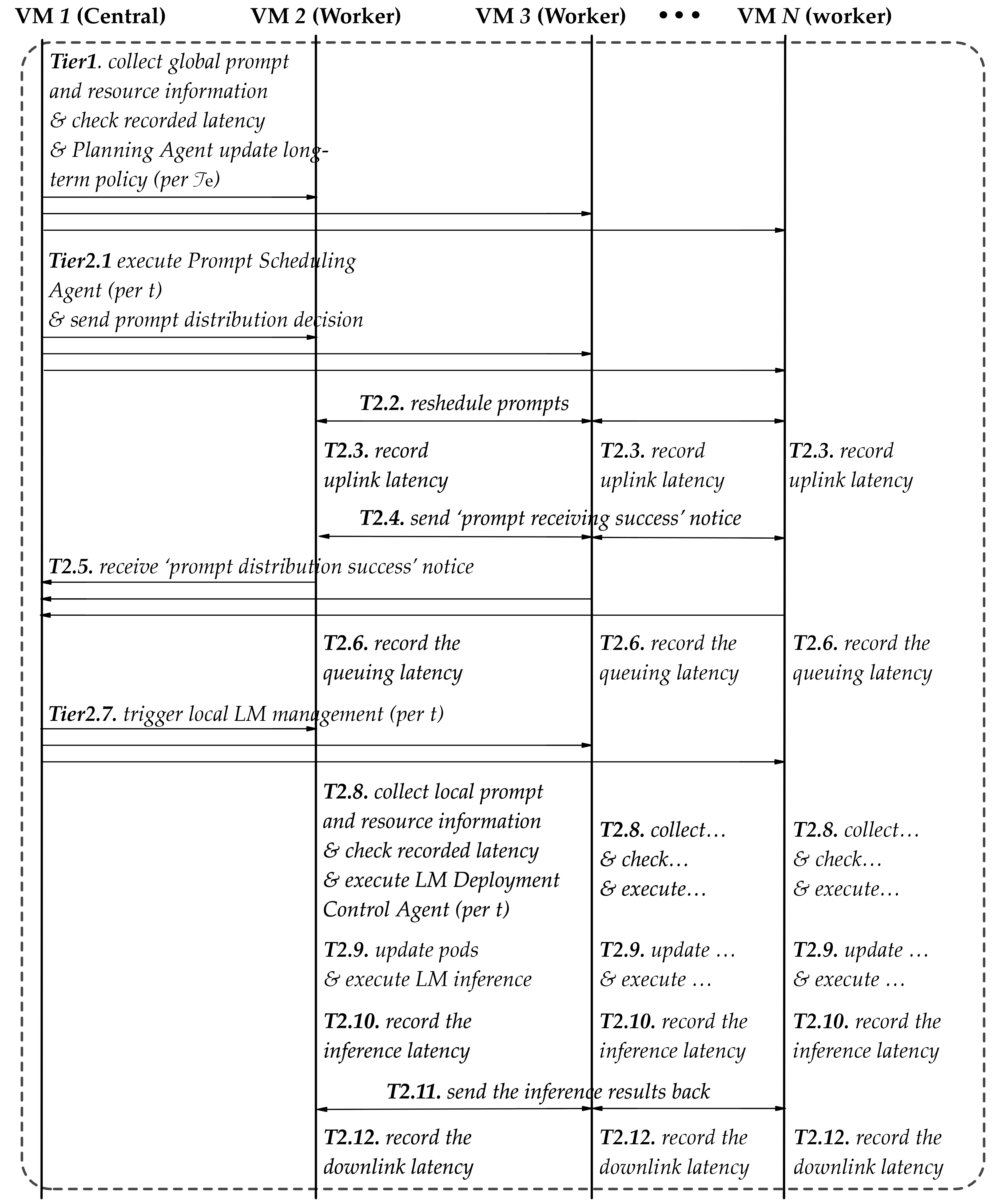}
    \caption{Sequence diagram of the runtime operation. Communications between MEC servers are achieved by Fast API.}
    \vspace{-0.2cm}
    \label{fig:flows}
\end{figure}

Building upon the aforementioned network setup, this section explores the runtime operation of the two-tiered solution. A detailed sequence diagram is presented in Figure~\ref{fig:flows}.
In Tier-1 (Long-term Planning), the central VM first collects global network state and executes the Planning Agent per $\mathcal{T}_{e}$. 

Once the macro-policy $(\mathcal{P}_e, \mathcal{R}_e)$ is generated, the system proceeds with short-term operations in each time slot $t$. At the beginning of the slot, the Prompt Scheduling Agent is executed based on the latest system state observed at the end of the previous slot and generates the slot-level routing map that defines $\{\mathcal{D}_q\}$ for incoming prompts (T2.1).
The resulting routing decisions are then disseminated to all MEC servers (T2.1). Each server then schedules its assigned prompts accordingly (T2.2). During the prompt scheduling process, uplink transmission latencies are recorded (T2.3). Upon receiving the dispatched prompt sets, the corresponding pods notify both the sending servers and the controller (T2.4, T2.5), and begin tracking the queuing time for each request (T2.6).

The execution of the LM Deployment Agent is triggered (T2.7) once all prompt routing decisions have been completed. This activates the LM Deployment Agent (T2.8), which reorganizes the resource allocation among LM instances based on the current queued workload and available resources. The new configuration is applied (T2.9), and prompts are processed by the scheduled LMs. Inference latency is recorded during this process (T2.10). After processing, the inference results are sent back to the originating servers (T2.11), and the downlink latency is recorded (T2.12).

Within these processes, each node concurrently handles both Input/Output (I/O)-bound tasks, such as sending and receiving files or commands, deploying Agentic AI actions, and saving monitoring data to the MongoDB database. While executing these activities, all I/O-bound activities are implemented as asynchronous operations using \textit{async} and \textit{httpx}. This approach prevents blocking calls that could lead to event loop deadlocks and allows the system to yield execution when awaiting network responses or file I/O, rather than stalling entirely.

\vspace{-0.3cm}
\section{Experiment Results}
\label{sec:results}

Performance measurement for AI-centric computing-continuum systems depends on the target service objective, workload dynamics, and deployment scale~\cite{donta2025performance}. Accordingly, we evaluate the proposed framework through multiple complementary views, including overall latency-fairness performance, per-LM service quality, service-type fairness over time, convergence behavior, workload-shift adaptation, and expanded-testbed scalability. The evaluation compares the proposed multi-agentic framework (MA) with the following baselines

\noindent \textit{1) Random scheduling \& random deployment (RR):}
Prompts are assigned randomly across servers, and each MEC node randomly activates LMs within local resource budgets, regardless of queue backlogs or prompt composition.

\noindent \textit{2) Random scheduling \& Lyapunov-based LM deployment (RL):} Requests are routed randomly across nodes, while each server independently solves 
\begin{equation}
    \quad \min_{\mathcal{G}} \;
    \mathbb{E}\!\left[ \lambda \bar{T}_{\mathrm{norm}} + (1-\lambda)\big(1 - F_{\mathrm{norm}}(\boldsymbol{\rho})\big)
    \mid \mathcal{D} \right],
    \label{eq:sub2}
\end{equation}
using a Lyapunov drift-plus-penalty (DPP) problem~\cite{neely2010stochastic} to determine which LMs to activate on CPU/GPU. 

Specifically, each server maintains a finite set of candidate deployment configurations, where each action specifies, for each LM type, whether it is inactive, deployed on CPU, or deployed on GPU. Actions that violate resource budgets or conflict with transient pod states are filtered out, yielding a feasible action set. At every slot $t$, the controller selects an action that maximizes the following score~\cite{neely2010stochastic}:
\begin{equation}
\sum_{i \in \mathcal{I}} Q_{ni}(t) \cdot \hat{\mu}_{ni}(a)
- V \cdot C_n(a,t),
\end{equation}
where $Q_{ni}(t)$ denotes the queue backlog of LM type $i$ at server $n$, $\hat{\mu}_{ni}(a)$ is a proxy for the service rate under configuration $a$, and $C_n(a,t)$ represents the per-slot deployment cost. The parameter $V$ controls the trade-off between backlog reduction and operational overhead.

The service-rate proxy $\hat{\mu}_{ni}(a)$ depends on the compute mode assigned to LM type $i$ and the residual resource headroom. We adopt a smoothed multiplicative model:
\begin{equation}
\hat{\mu}_{ni}(a) =
\begin{cases}
\alpha_i^{\mathrm{cpu}}\, f\big(\eta_{n}^{\mathrm{cpu}}(t)\big)\, f\big(\eta_{n}^{\mathrm{mem}}(t)\big), 
    & i \in \mathcal{I}_{n,\mathrm{cpu}}^{\text{act}}(a),\\[3pt]
\alpha_i^{\mathrm{gpu}}\, f\big(\eta_{n}^{\mathrm{gpu}}(t)\big), 
    & i \in \mathcal{I}_{n,\mathrm{gpu}}^{\text{act}}(a),\\[3pt]
0, & \text{otherwise},
\end{cases}
\end{equation}
where $\eta_n^{\cdot}(t)$ denotes normalized residual CPU, memory, and GPU capacities, $f(\cdot)$ is a linear smoothing function, and $\alpha_i^{\mathrm{cpu}}$ and $\alpha_i^{\mathrm{gpu}}$ denote the CPU-mode and GPU-mode service-rate weights of LM type $i$, respectively.

The per-slot cost $C_n(a,t)$ consists of two components. The GPU usage cost is defined as
\begin{equation}
C_n^{\mathrm{gpu}}(a,t) = \pi_{n}^{\mathrm{gpu}}(t) \cdot N_{\mathrm{GPU}}(a),
\end{equation}
where $N_{\mathrm{GPU}}(a)$ denotes the number of GPU-backed LM instances under action $a$. The GPU shadow price $\pi_{n}^{\mathrm{gpu}}(t)$ increases with GPU saturation and the fraction of image-type requests:
\begin{equation}
\pi_{n}^{\mathrm{gpu}}(t) = p_0 + p_1 \big(1 - \eta_n^{\mathrm{gpu}}(t)\big) + p_2 \, \phi_n^{\mathrm{img}}(t),
\end{equation}
where $\phi_n^{\mathrm{img}}(t)$ denotes the local image workload ratio. 

To discourage frequent activation and deactivation of large models, we introduce a churn penalty:
\begin{equation}
{
\begin{aligned}
\Delta_{n,i}^{\mathrm{act}}(a,t)
\triangleq{}&
\Bigl|
\mathbbm{1}\!\left[i\in\mathcal{I}_n^{\mathrm{act}}(a)\right] \\
&-
\mathbbm{1}\!\left[i\in\mathcal{I}_n^{\mathrm{act}}(t-1)\right]
\Bigr|, \\
C_n^{\mathrm{churn}}(a,t)
={}&
\lambda_{\mathrm{churn}}
\sum_{i\in\mathcal{I}}
\Delta_{n,i}^{\mathrm{act}}(a,t)
\bigl(1+\kappa Q_{ni}(t)\bigr).
\end{aligned}
}
\label{eq:churn_cost}
\end{equation}
The total cost is given by $C_n(a,t) = C_n^{\mathrm{gpu}}(a,t) + C_n^{\mathrm{churn}}(a,t)$.
Therefore, the DPP score above contains two sets of tunable quantities: the service-rate weights $\alpha_i^{m}$ and the cost coefficients $(p_1,p_2,\lambda_{\mathrm{churn}},\kappa)$. To determine these quantities, the Lyapunov-based deployment algorithm  is parameterized through a pre-benchmark calibration trace and then kept fixed during benchmark evaluation. The calibration trace uses randomized feasible routing to expose the deployment algorithm  to heterogeneous prompt arrivals while collecting synchronized deployment-mode, routing, backlog, and completed-request latency telemetry, starting from an unprofiled prior in which all feasible LM/mode service-rate weights are initialized to one and infeasible modes to zero.

First, the service-rate weights in the service-rate proxy above are estimated from measured completions. For LM type $i$ and execution mode $m\in\{\mathrm{cpu},\mathrm{gpu}\}$, let $\Gamma_i^m$ be the number of completed requests attributed to mode $m$ during a calibration window $\mathcal{W}_{\mathrm{cal}}$, where the serving node and mode of each completed request are recovered by joining its origin-side latency record, the routing decision at the arrival slot, and the destination's deployment snapshot at the estimated completion time. The empirical service rate is
\begin{equation}
\hat{\mu}_{i}^{m}
=
\frac{\Gamma_i^m}
{\sum_{n\in\mathcal{N}}\sum_{t\in\mathcal{W}_{\mathrm{cal}}}
\mathbbm{1}\{r_{ni}(t)=m,\,Q_{ni}(t)>0\}\Delta t},
\end{equation}
where $r_{ni}(t)\in\{\mathrm{off},\mathrm{cpu},\mathrm{gpu}\}$ is the deployment mode of LM $i$ on node $n$, $\Delta t$ is the slot length, and the denominator counts slot-level active and backlogged server time in mode $m$. The weights $\alpha_i^m$ are then obtained by normalizing $\hat{\mu}_i^m$ by the smallest positive estimated rate. LM-mode pairs with insufficient completions or active-backlogged time retain the uniform feasible-mode prior and are recorded in the calibration log, while pairs excluded by the deployment feasibility constraints are assigned zero weight.

The coefficient search uses the following calibration objective:
\begin{equation}
J_{\mathrm{cal}}(x)
=
\frac{\bar{L}_{\mathrm{cal}}(x)}{\tau}
+
\left(1-F_{\mathrm{cal}}(x)\right)
+
\bar{C}_{\mathrm{flip}}(x),
\label{eq:lyap-calibration-objective}
\end{equation}
where $x=\log\theta$, $\theta=(p_1,p_2,\lambda_{\mathrm{churn}},\kappa)$, $\bar{L}_{\mathrm{cal}}$ is the average of the per-LM success-only mean latencies measured over the calibration window, with $\tau$ assigned to an LM type that has no successful completion in that finite window, and $F_{\mathrm{cal}}$ is Jain's service-type fairness over the same window. The flip term $\bar{C}_{\mathrm{flip}}$ is the average per-slot fraction of LM deployment-mode changes among $\{\mathrm{off},\mathrm{cpu},\mathrm{gpu}\}$ across nodes and LM types in the evaluation window. It is used only for calibration-window evaluation, while the runtime churn term in Eq.~\eqref{eq:churn_cost} penalizes activation-state changes inside the DPP score. This calibration objective is used only to choose stable DPP coefficients, while the reported benchmark results continue to use latency and fairness metrics defined in Section~\ref{sec:problem}.

\begin{figure}[t]
    \centering
    \setlength{\abovecaptionskip}{0cm}
    \subfigure[Calibrated parameters]{
        \label{fig:lyapunov-calibration-params}
        \includegraphics[width=0.478\linewidth]{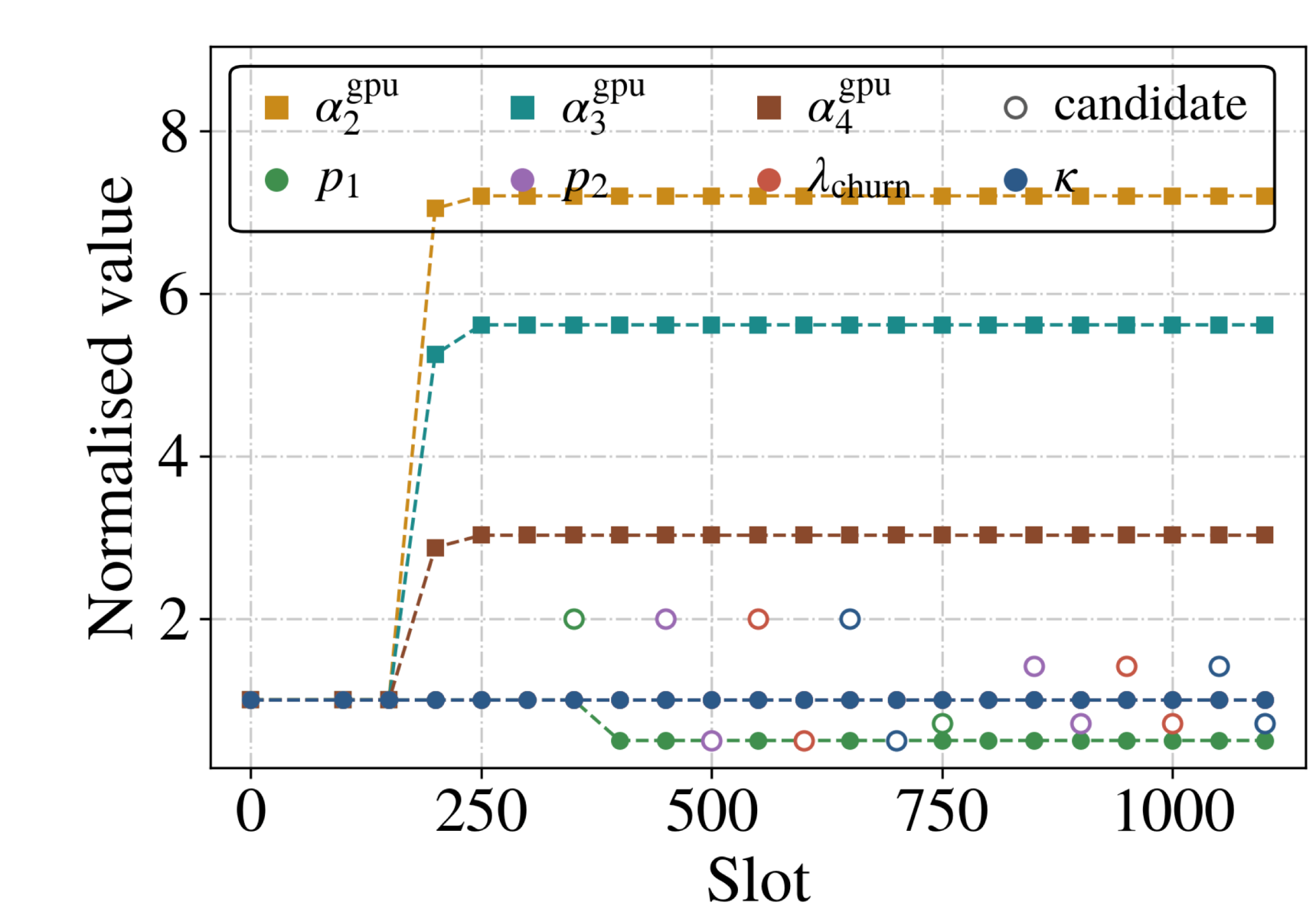}}
    \subfigure[Windowed objective]{
        \label{fig:lyapunov-calibration-cost}
        \includegraphics[width=0.478\linewidth]{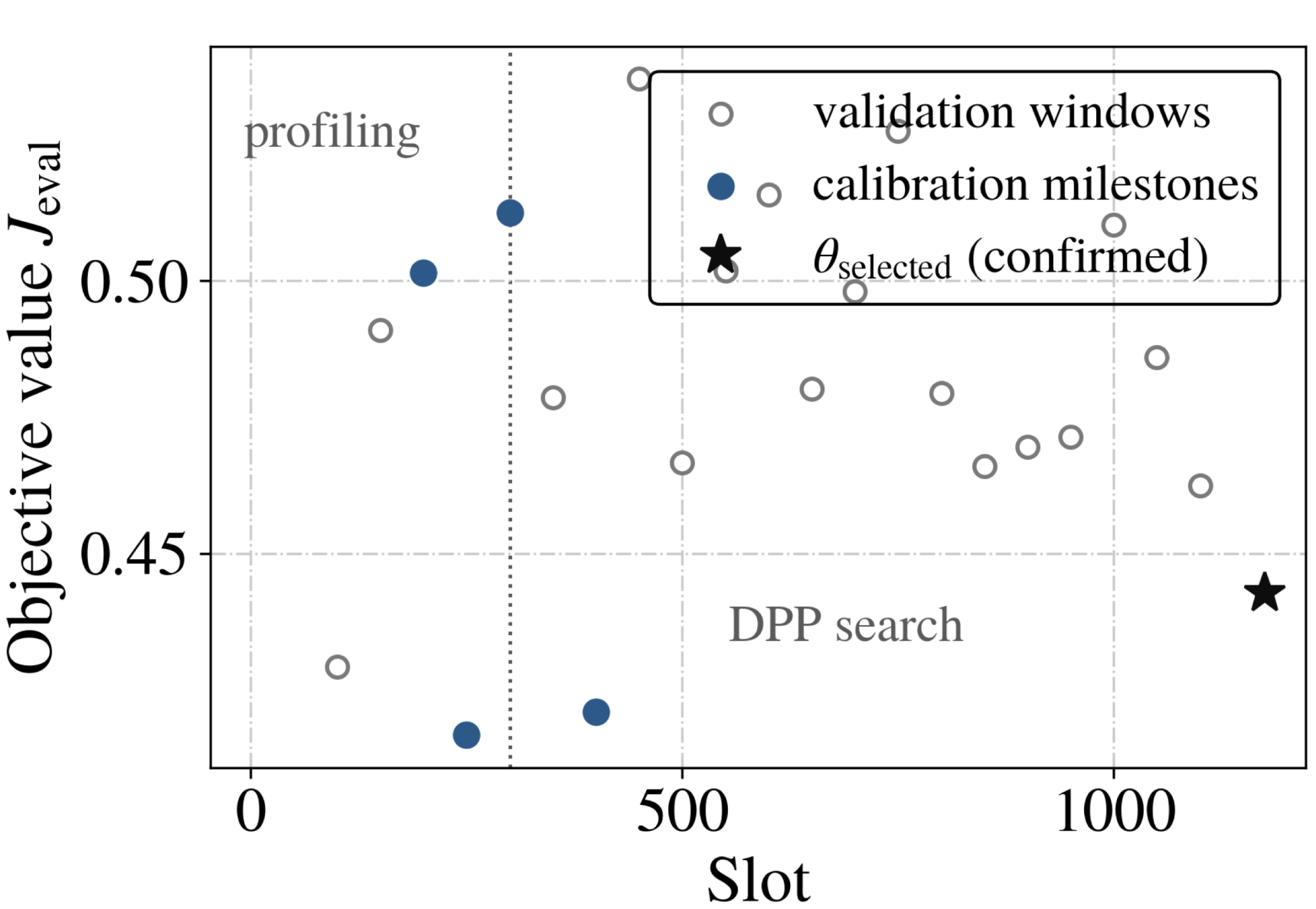}}
    \caption{Pre-benchmark calibration of the Lyapunov-based deployment algorithm. (a) Trajectory of the profiled service-rate weights and margin-accepted DPP coefficients. (b) Raw validation-window objective and objective of the active calibrated configuration.}
    \label{fig:lyapunov-calibration}
    \vspace{0cm}
\end{figure}

Second, after the service-rate weights are selected, the DPP coefficients are calibrated by a deterministic log-space coordinate search. Starting from a generic positive prior, the search perturbs one coefficient at a time in a fixed coordinate order, evaluating a multiplicative increase and decrease of the current value over a predefined sequence of search radii. The current point and the two candidates are compared on validation windows measured at the actual coefficient values, and a candidate is accepted only when it improves $J_{\mathrm{cal}}$ by more than a fixed acceptance margin $\epsilon_{\mathrm{acc}}$; otherwise, the current value is retained. The margin is fixed before the calibration run and prevents coefficient updates from being selected on small objective fluctuations, whose variation across repeated cold-start windows is recorded as a noise diagnostic.
The selected DPP coefficients are the final margin-accepted values, and separate confirmation windows are used only to report the final calibration objective. Figure~\ref{fig:lyapunov-calibration} reports the resulting parameter trajectory and validation-window calibration objective.

The finalized Lyapunov parameter set used by all Lyapunov-based baselines is $(p_0,p_1,p_2,\lambda_{\mathrm{churn}},\kappa)=(0,0.15,0.30,0.10,1.00)$, with service-rate weights $\boldsymbol{\alpha}^{\mathrm{cpu}}=(3.68,1.00,0,0)$ and $\boldsymbol{\alpha}^{\mathrm{gpu}}=(5.80,7.20,5.61,3.02)$ for LM1--LM4, respectively. The scalar $V$ is set to $1.0$, so the relative weighting between backlog reward and deployment cost is determined by the calibrated coefficients $(p_1,p_2,\lambda_{\mathrm{churn}},\kappa)$. These values are fixed during benchmark execution. The resulting algorithm defines  a queue-aware deployment policy. Backlog terms prioritize LM types with accumulated requests, while GPU-price and churn terms moderate GPU contention and repeated activation changes.

\noindent \textit{3) Agentic scheduling \& Lyapunov-based deployment (MAL):} This setting uses our proposed Agentic AI-based Planning Agent and Prompt Scheduling Agent for routing decisions, while still relying on the Lyapunov-based deployment algorithm  for local LM deployment. 

\noindent \textit{4) Average scheduling \& Lyapunov-based deployment (AL):} Requests are uniformly distributed across all feasible nodes (i.e., those capable of supporting the LM type), and each server uses the Lyapunov-based deployment algorithm  to decide resource allocation. 

\noindent \textit{5) Random scheduling \& full activation (RF):} Prompts are randomly routed across nodes, and each server attempts to activate as many LMs as possible within its CPU/GPU constraints, without dynamic reconfiguration. 

\noindent \textit{6) Local processing \& Lyapunov-based deployment (LL):} All prompts are processed on their arrival node without inter-node dispatching, and each server runs the Lyapunov-based deployment algorithm  to manage local LMs. 

\noindent \textit{7) PerLLM:} This setting replaces the Prompt Scheduling Agent with the upper-confidence-bound (UCB) routing bandit of Yang et al.~\cite{yang2024perllm}, while keeping the same Lyapunov-based deployment algorithm  for local LM deployment. We instantiate one bandit for each origin-node/LM-type pair, with feasible destination nodes as arms and a latency-derived reward $1-\min(d/\tau,1)$, where $d$ is the observed end-to-end delay of the routed request and $\tau$ is the service deadline. Since the reward is available only after the request completes or times out, updates are applied through a pending-reward queue, so PerLLM is evaluated under the same delayed-feedback deployment pipeline as the proposed framework.

\noindent \textit{8) MILP-based joint routing and deployment (MILP):} This solution uses an MILP to jointly determine prompt routing and LM activation at each slot. It observes the current request arrivals, realized per-node queues, and current deployment states, and selects a binary destination for each nonempty origin-node/LM-type request batch and a feasible GPU-backed LM activation configuration for each worker. The static MILP input data are obtained through pre-benchmark physical profiling of execution times across LM types, request batch sizes, GPU-node classes, and co-located LM configurations, together with deployment-transition and pairwise transfer delays. These profiled inputs remain fixed during evaluation, whereas the queue and deployment states are refreshed before each optimization step. Because the objective in Eq.~\eqref{eq:obj} contains the nonlinear Jain term and realized end-to-end latency, the MILP uses a linear surrogate comprising estimated end-to-end request delay, the mean pairwise absolute difference between modeled per-LM service ratios, and a deployment-transition penalty. The delay estimate combines measured transfer delay, observed queue backlog, profiled inference time, and applicable cold-start delay. 

Importantly, for all Non-local Processing, requests for LM3 and LM4 are never routed to VM2 as it doesn't have GPU resources.

\vspace{-0.45cm}
\subsection{Latency-Fairness Performance}
\begin{figure}[t]
    \centering
    \setlength{\abovecaptionskip}{0.2cm}
    \includegraphics[width=0.99\linewidth]{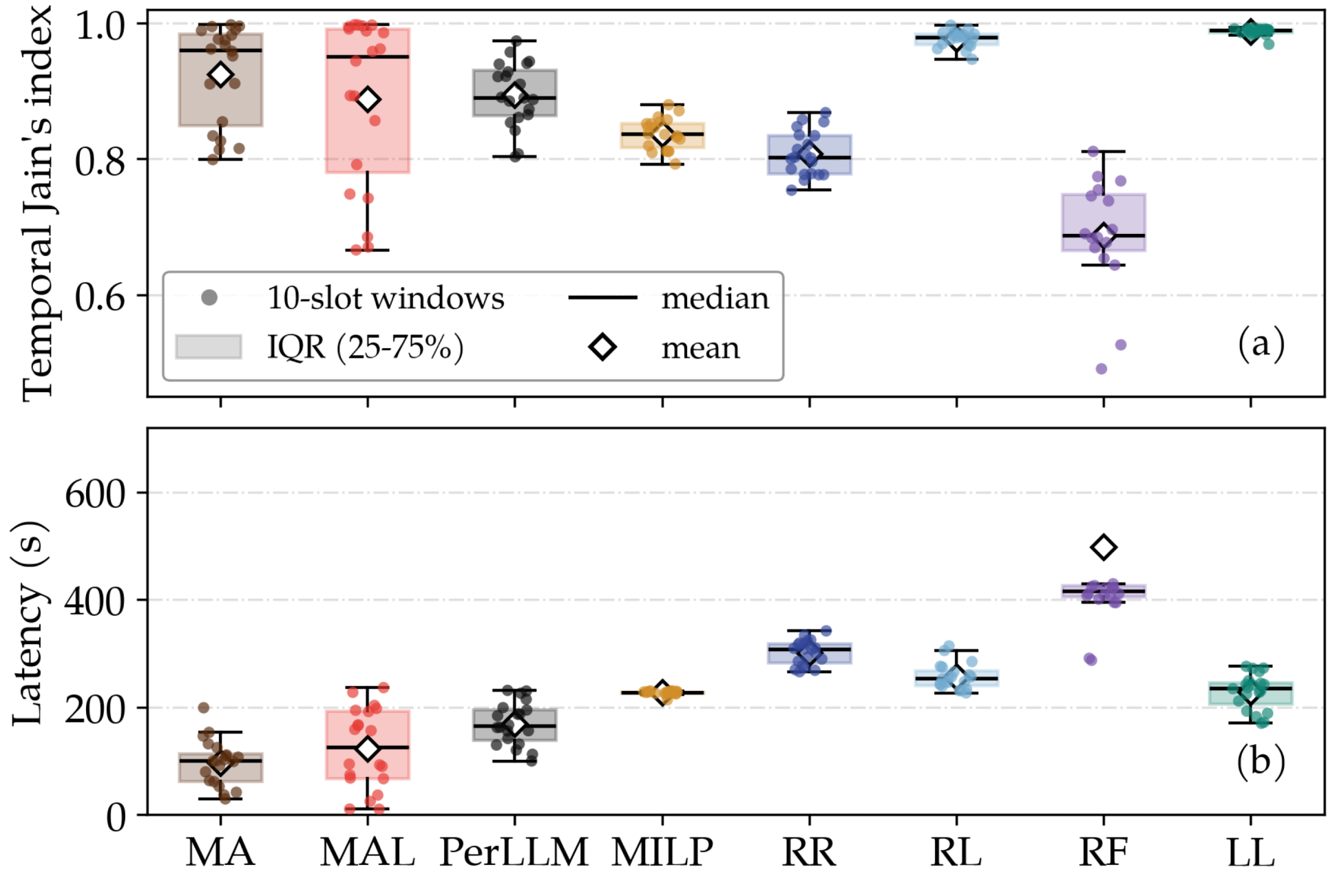}
    \vspace{-0.6cm}
    \caption{Comparison of different scheduling and deployment strategies in terms of service-type fairness and latency. (a) Jain's fairness index over LM service-type success ratios. (b) Latency over the same windows.}
    \vspace{-0.3cm}
    \label{fig:temporal}
\end{figure}

To evaluate latency and service-type fairness over time, Fig.~\ref{fig:temporal} compares all scheduling and deployment strategies over 200-slot evaluation segments. Each segment is partitioned into fixed-length time windows. For each window, Fig.~\ref{fig:temporal}(a) reports Jain's fairness index computed from the per-LM success ratios, while Fig.~\ref{fig:temporal}(b) reports latency over the same windows. The boxplots summarize the resulting distributions within each evaluation segment.
The baseline distributions reveal three patterns. First, resource activation without coordinated routing leads to poor latency-fairness trade-offs: RF has the highest mean latency of $497.8$\,s and the lowest mean Jain's index of $0.688$, while RR improves latency but remains limited in fairness, with a mean Jain's index of $0.808$. Second, local LM deployment can improve service balance, but latency remains high without coordinated prompt routing. RL and LL maintain high mean Jain's indices of $0.977$ and $0.988$, respectively, yet their mean latencies across the windows remain $256.8$\,s and $228.6$\,s. Third, MILP jointly optimizes routing and deployment but reaches a mean latency of $226.6$\,s and a mean Jain's index of $0.837$. PerLLM improves these values to $169.3$\,s and $0.895$, respectively. This improvement is consistent with its use of realized rewards in UCB routing and observed queues in Lyapunov-based deployment, which partly mitigates the profile mismatches affecting MILP. MAL further lowers latency to $123.0$\,s, but its separate deployment mechanism results in a wider fairness distribution and a lower mean Jain's index of $0.888$. MA achieves the best latency--fairness balance, with a mean latency of $96.7$\,s and a mean Jain's index of $0.925$, reducing latency by $42.9\%$, $21.4\%$, $57.3\%$, and $80.6\%$ relative to PerLLM, MAL, MILP, and RF, respectively. These results suggest that MA benefits from vertical coordination across traffic planning, prompt dispatching, and local LM activation.

\begin{figure}[t]
    \centering
    \setlength{\abovecaptionskip}{0.2cm}
    \includegraphics[width=0.99\linewidth]{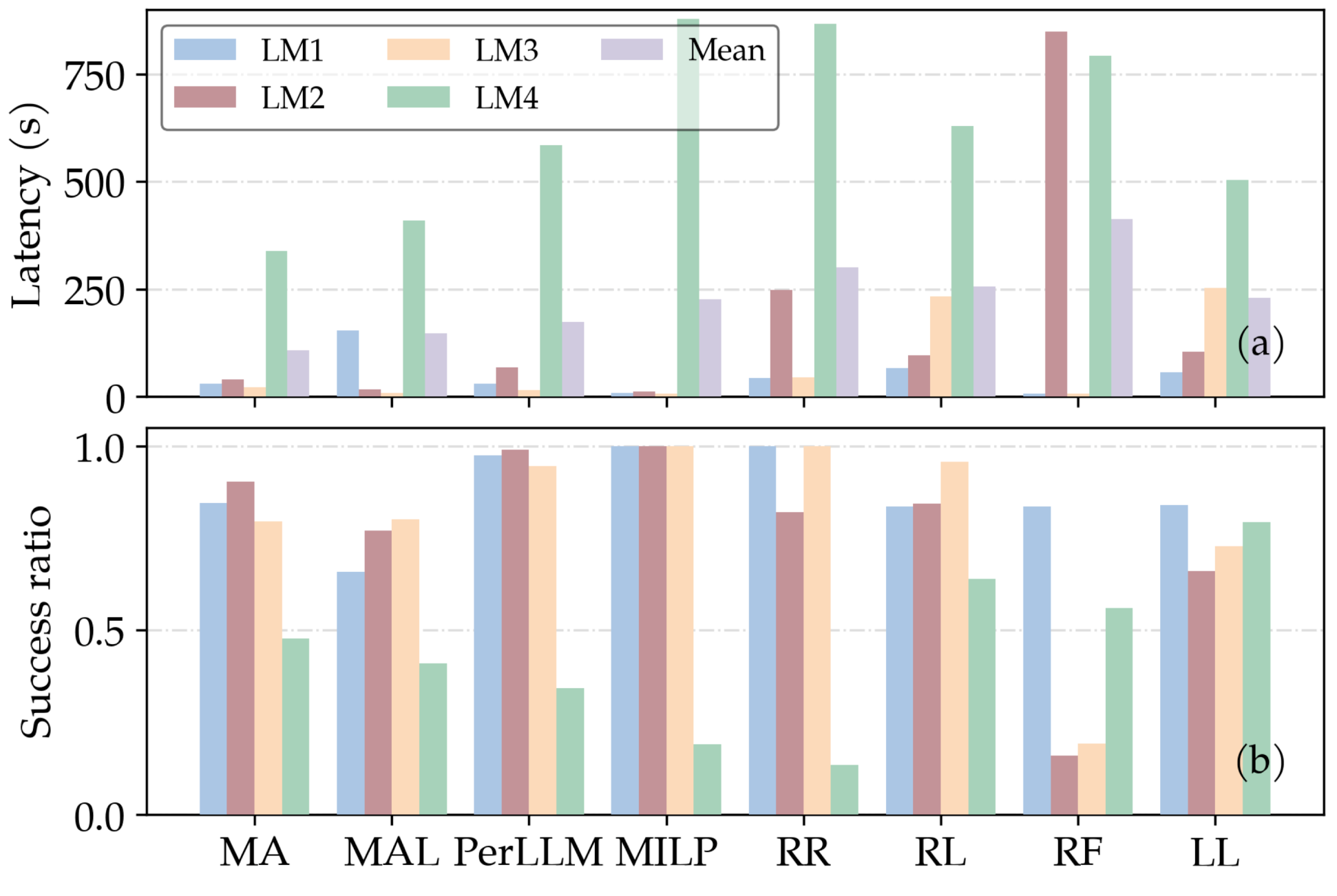}
    \vspace{-0.6cm}
    \caption{Latency and success ratio of each LM. (a) End-to-end latency for each LM and the overall mean. (b) Success ratio.} 
    \vspace{-0.2cm}
    \label{fig:detailed}
\end{figure}

To further explain the aggregate differences, Figure~\ref{fig:detailed} reports the per-LM end-to-end latency and service success ratio under each strategy. LM4 is the main latency bottleneck, showing the highest latency under MA, MAL, PerLLM, MILP, RR, RL, and LL. MILP exhibits the sharpest imbalance, with success ratios of $1.000$ for LM1--LM3 and mean latencies of $8.0$, $12.6$, and $7.2$\,s, respectively, while LM4 has a success ratio of $0.192$ and a mean latency of $877.7$\,s. This imbalance reflects MILP's sensitivity to deviations between its fixed execution, transfer, and transition profiles and the realized service times of long-running LM4 requests. PerLLM and MAL partly mitigate this limitation through runtime feedback, yielding LM4 latencies of about $584$\,s and $410$\,s and success ratios of $0.342$ and $0.411$, respectively. MA further coordinates routing and deployment using realized latency and per-LM success outcomes. It keeps the latencies of LM1--LM3 below $40$\,s and reduces LM4 latency to about $339$\,s with a success ratio of $0.479$, resulting in the lowest overall mean latency.

\vspace{-0.45cm}
\subsection{Cross-layer Coordination and Agent Reliability}
Runtime alignment between routing and deployment decisions is measured through the cross-node inactive-target conflicts reported in Table~\ref{tab:cross-layer-conflict}. A conflict occurs when a routed prompt reaches a destination node where the requested LM is not active at the arrival time. We compute this metric by joining each cross-node prompt-arrival record with the destination node's active-LM snapshot and the latency-component record stored in the MongoDB database for the same request.
Queue-delay statistics are estimated from the telemetry-based residual of the end-to-end delay after subtracting the recorded transmission and inference components, consistent with the decomposition of Eq.~\ref{eq:latency_sum}. For timed-out requests, the timeout deadline $\tau$ is retained as the delay value so that unresolved conflicts are included in the conflict-cost statistics.

\begin{table}[!t]
\centering
\caption{Cross-node inactive-target conflict analysis.}
\label{tab:cross-layer-conflict}
\scriptsize
\setlength{\tabcolsep}{2pt}
\renewcommand{\arraystretch}{1.08}
\begin{tabularx}{\columnwidth}{@{}
l
>{\centering\arraybackslash}X
>{\centering\arraybackslash}X
>{\centering\arraybackslash}X
>{\centering\arraybackslash}X
@{}}
\toprule
\textbf{Method} &
\makecell{\textbf{Inactive-target}\\\textbf{ratio}} &
\makecell{\textbf{Conflict queue}\\\textbf{delay}} &
\makecell{\textbf{Non-conf. queue}\\\textbf{delay}} &
\makecell{\textbf{Conflict timeout}\\\textbf{ratio}} \\
\midrule
MAL &
6.36\% & 679.9\,s & 127.1\,s & 44.46\% \\
PerLLM &
6.29\% & 480.9\,s & 216.7\,s & 37.45\% \\
MA &
7.25\% & 302.3\,s & 98.4\,s & 17.12\% \\
\bottomrule
\end{tabularx}
\end{table}

The conflict statistics show that similar mismatch frequencies can lead to very different conflict costs. MAL, PerLLM, and MA have comparable inactive-target ratios, ranging from $6.29\%$ to $7.25\%$. The methods differ mainly in the cost incurred after a mismatch occurs. Under MAL, a conflict leads to a mean queue delay of $679.9$\,s and a timeout ratio of $44.46\%$; PerLLM reduces the conflict delay to $480.9$\,s but still has a conflict timeout ratio of $37.45\%$. MA lowers the conflict queue delay to $302.3$\,s and the conflict timeout ratio to $17.12\%$, while also reducing the non-conflict queue delay to $98.4$\,s. This pattern matches the design difference among the three methods. In MA, prompt scheduling and local deployment are both generated within the same agentic hierarchy and conditioned on shared runtime observations, so a node that receives backlogged requests for an inactive LM can be adjusted in both subsequent routing and local activation. In MAL and PerLLM, routing decisions and Lyapunov-based deployment decisions are produced by separate mechanisms: the router may continue to send requests according to its traffic policy, while the deployment algorithm activates LMs according to local queue and cost terms. This looser coupling makes a temporarily inactive-target event more likely to persist into long queues or timeout records.

The reliability of LLM-agent decisions is evaluated by recording whether each generated output is parsed, validated, accepted, or replaced by a fallback action. The audit is restricted to the control period corresponding to $65$ Global Planning calls, or nominally $3{,}250$ slots. All $65$ calls produce valid macro-policy outputs. Six terminal slots do not contain completed Prompt Scheduling records, leaving $3{,}244$ recorded calls, all of which produce valid routing outputs. Among these slots, $3{,}234$ have complete six-worker deployment-state snapshots, producing $19{,}404$ valid node-slot placement outputs. No parsing or validation failures occur within this aligned audit window. The only runtime exception is one Prompt Scheduling API failure, triggering the deterministic fallback that keeps traffic on the origin node's local-routing map. The latency and fairness measurements in Figures~\ref{fig:temporal} and~\ref{fig:detailed} include this fallback event, so the reported results reflect the executed recovery path. This reliability audit measures interface-level failures, while decision quality is evaluated through the closed-loop performance results. Valid but suboptimal routing or deployment choices are executed in the closed loop, and their effects are reflected in the reported latency, service-type fairness, per-LM success ratios, queue evolution, and cross-layer conflict results.

\begin{table}[!t]
\centering
\caption{MA deployment reconfiguration frequency. }
\label{tab:reconfiguration-frequency}
\scriptsize
\setlength{\tabcolsep}{1.8pt}
\renewcommand{\arraystretch}{1.08}
\begin{tabularx}{\columnwidth}{@{}
l
>{\centering\arraybackslash}X
>{\centering\arraybackslash}X
@{}}
\toprule
\textbf{LM type} &
\makecell{\textbf{Activation events}\\\textbf{per 100 node-slot}\\\textbf{decisions}} &
\makecell{\textbf{Deactivation events}\\\textbf{per 100 node-slot}\\\textbf{decisions}} \\
\midrule
LM1 & $2.28$ & $2.26$ \\
LM2 & $1.61$ & $1.62$ \\
LM3 & $2.32$ & $2.34$ \\
LM4 & $0.61$ & $0.58$ \\
\midrule
All LMs & $6.82$ & $6.81$ \\
\bottomrule
\end{tabularx}
\vspace{-0.2cm}
\end{table}

To quantify pod reconfiguration overhead, Table~\ref{tab:reconfiguration-frequency} further reports the activation and deactivation frequencies of MA over the audit window. An inactive-to-active change is counted as an activation event, and an active-to-inactive change is counted as a deactivation event. Because multiple LM states can change within the same worker-slot transition, Table~\ref{tab:reconfiguration-frequency} reports both per-LM event frequencies and the worker-level active-set change rate. Across $19{,}404$ local deployment decisions, MA produces $1{,}323$ activation events and $1{,}321$ deactivation events, corresponding to $6.82$ and $6.81$ events per $100$ decisions, respectively. The displayed per-LM rates may not sum exactly to the all-LM rates because each value is rounded independently to two decimal places. The active LM set changes in $11.89\%$ of the $19{,}374$ consecutive worker-slot transitions and remains unchanged in the other $88.11\%$. Therefore, pod activation and deactivation overhead is incurred in only a minority of control transitions. The pod-startup, termination, pending-state, and interrupted-execution effects of these events are included in the measured queueing delays, timeout outcomes, and service success ratios.

\vspace{-0.4cm}
\subsection{Learning Dynamics}
\begin{figure}[t]
    \centering
    \setlength{\abovecaptionskip}{0cm}
    \subfigure[MA]{
        \label{fig:rl}
        \includegraphics[width=0.478\linewidth]{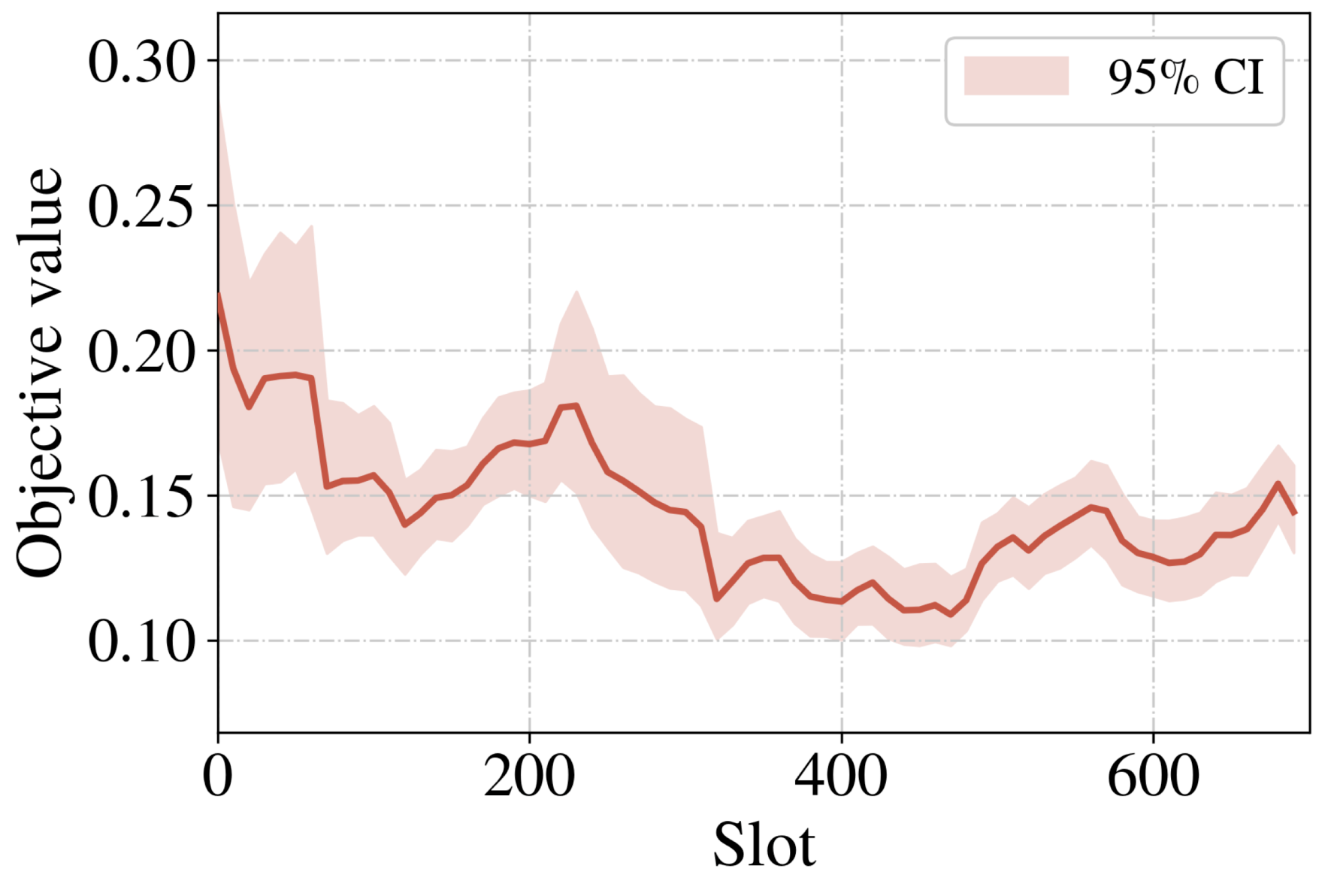}}
    \subfigure[PerLLM]{
        \label{fig:perllm-rl}
        \includegraphics[width=0.478\linewidth]{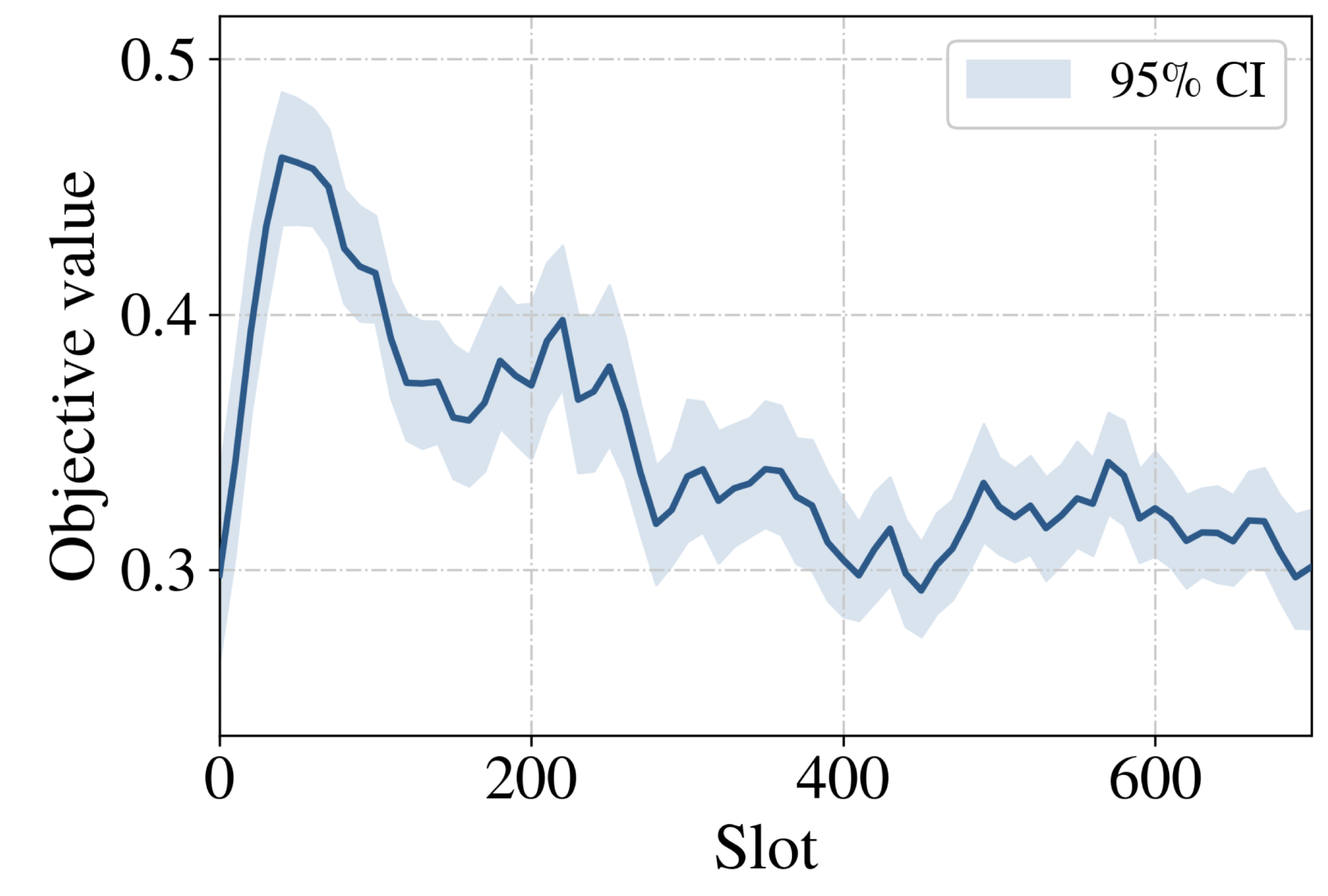}}
    \caption{Learning dynamics. Both panels report rolling-window objective values with bootstrap 95\% confidence intervals. MA updates occur once per epoch.}
    \label{fig:learning-dynamics}
    \vspace{-0.1cm}
\end{figure}

To assess learning dynamics, Fig.~\ref{fig:learning-dynamics} reports rolling-window objective trajectories for MA and PerLLM under delayed feedback. Both panels use the same latency-fairness objective, and the shaded regions show bootstrap 95\% confidence intervals over the requests within each window. MA makes macro-policy updates once per epoch, so its curve reflects the effect of a small number of planning updates propagated through slot-level routing and deployment decisions. The objective drops after the initial planning stage and remains in a lower range, indicating that the episodic-memory planner quickly identifies useful traffic and deployment patterns. PerLLM continuously updates its UCB routing decisions at the slot level. Its objective decreases more gradually after the early exploration period and remains higher than MA over most of the evaluated slots, showing that online bandit learning requires more request-level feedback before reaching a stable operating point. This comparison complements the steady-state results by showing both convergence behavior and the different feedback timescales of the two learning mechanisms.

\begin{figure}[t]
    \centering
    \setlength{\abovecaptionskip}{0cm}
    \includegraphics[width=1\linewidth]{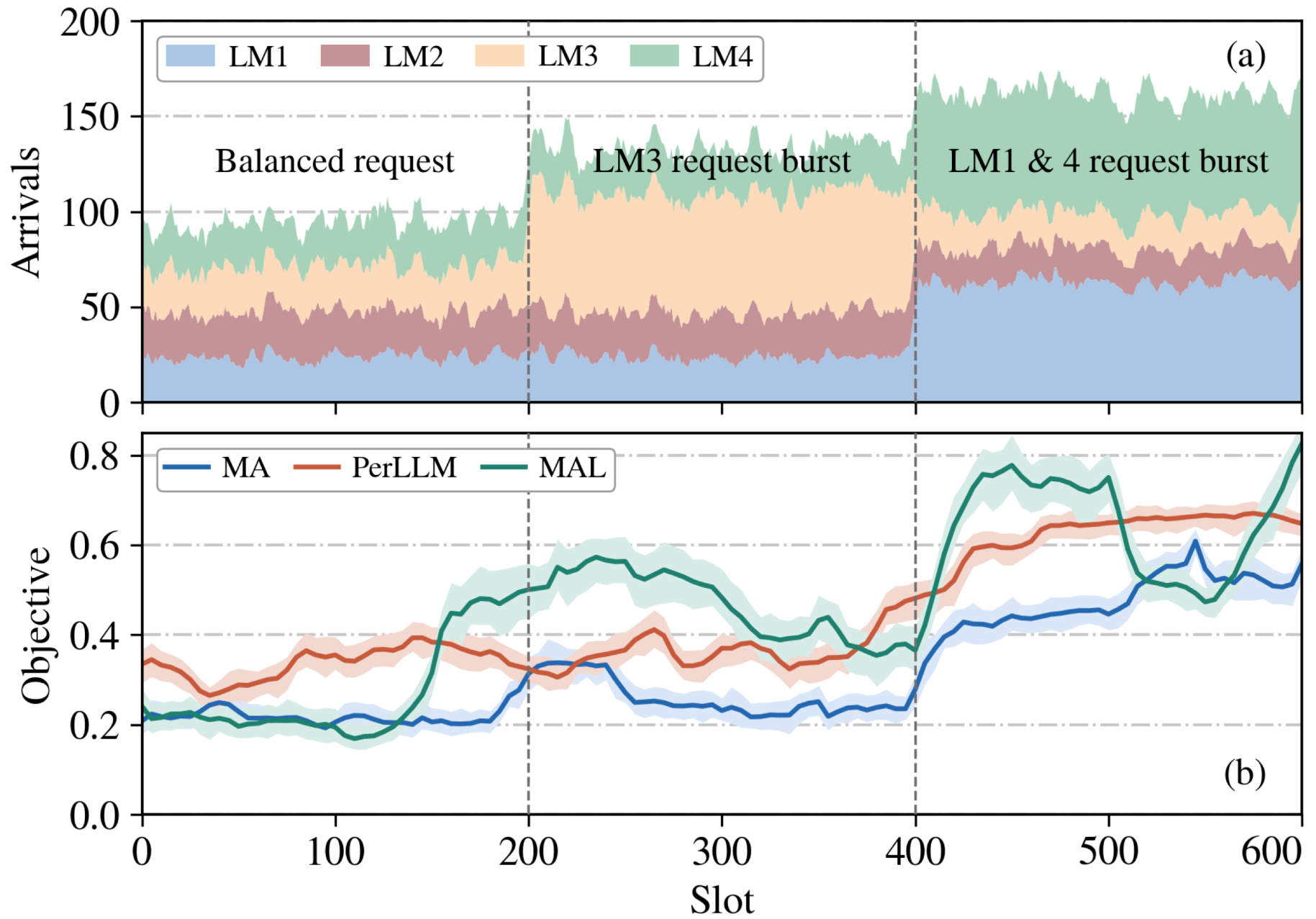}
    \caption{Heterogeneous bursty-arrival stress test. (a) Request arrivals by LM type. (b) Rolling latency--fairness objective with bootstrap 95\% confidence intervals.}
    \vspace{-0.3cm}
    \label{fig:bursty}
\end{figure}

We also implemented a hierarchical DRL baseline that mirrors the two-tier structure of our framework. The upper tier addresses inter-server prompt scheduling via Multi-Agent Proximal Policy Optimization (MAPPO): each agent observes local prompt queue lengths, active LM configurations, and a summary of global resource availability, and selects a destination server for each prompt type. Eq.~\eqref{eq:sub2} on the second tier is solved by one DQN agent per server, which observes local prompt queues and residual computational resources and selects a full LM deployment configuration, including the activated LM types and their compute modes, from a pre-enumerated discrete set. Both tiers are trained cooperatively with a shared delayed reward, issued only after all requests arriving in slot $t$ are completed and defined directly from the objective in Eq.~\eqref{eq:obj}: 
\begin{equation}
r_t = 1 - \left[ \lambda \bar{T}_{\mathrm{norm}}^{(t)} + (1-\lambda)\big(1 - F_{\mathrm{norm}}^{(t)}\big) \right],
\end{equation}
where $\bar{T}_{\mathrm{norm}}^{(t)}$ and $F_{\mathrm{norm}}^{(t)}$ are computed over completed requests from slot $t$.

However, despite extensive experimentation with hyperparameters, this baseline failed to converge to a stable policy. This may be attributed to several challenges. First, the two layers induce mutual non-stationarity since the deployment layer must adapt to an evolving routing distribution, while the scheduling layer learns against deployment patterns that are themselves still changing. Second, rewards are sparse and delayed, as a request receives a definitive success/failure signal only upon completion or expiration of the $\tau=900$\,s tolerance window. Third, the control timescales are coupled because each epoch-level upper-tier transition aggregates multiple slot-level routing and deployment decisions made while the lower-tier policies are still exploring. Such aggregation exposes the upper-tier reward to substantial delay, variance, and non-stationarity induced by the evolving lower-layer policies.

Nevertheless, failure of convergence is itself instructive. It demonstrates that traditional learning-based controllers often suffer from poor generalization and require extensive training.
Under an optimistic assumption that the DRL baseline converges within $20{,}000$ training steps, this would still require $600{,}000$ seconds ($166.67$ hours) of interaction time at 30-second slot intervals.
In contrast, the Agentic AI framework achieves stable performance within $6-15$ macro epochs (approximately $2.5-6.25$ hours). This stark contrast underscores the practicality and generalization advantages of LLM–powered agents in complex, dynamic systems.

\begin{table}[t]
\centering
\caption{Expanded-testbed scalability under the same per-node traffic-generation pattern.}
\label{tab:scalability}
\scriptsize
\setlength{\tabcolsep}{2pt}
\renewcommand{\arraystretch}{1.08}
\begin{tabularx}{\columnwidth}{@{}
l
>{\centering\arraybackslash}X
>{\centering\arraybackslash}X
>{\centering\arraybackslash}X
>{\centering\arraybackslash}X
>{\centering\arraybackslash}X
@{}}
\toprule
\textbf{Setting} &
\makecell{\textbf{Worker}\\\textbf{agents}} &
\makecell{\textbf{Avg.}\\\textbf{load/slot}} &
\makecell{\textbf{Lat.}\\\textbf{(s)}} &
\makecell{\textbf{Jain}\\\textbf{index}} &
\makecell{\textbf{Obj.}\\\textbf{value}} \\
\midrule
Original testbed &
$6$ &
$\approx 96$ &
$96.7$ &
$0.925$ &
$0.182$ \\
Expanded testbed &
$8$ &
$\approx 128$ &
$132.1$ &
$0.932$ &
$0.215$ \\
\bottomrule
\end{tabularx}
\end{table}

\subsection{Generalization and Scalability}
\vspace{-0.1cm}
We evaluate traffic generalization under sustained demand shifts using the bursty-arrival trace in Fig.~\ref{fig:bursty}(a). 
The trace preserves the same physical testbed and agent logic while changing the request arrival pattern across three operating regimes: 
a balanced mixed-arrival phase, an LM3 request burst, and an LM1/LM4 request burst. At each slot, arrivals are generated per origin node and LM type. The balanced phase uses the same normal arrival range for all LM types. The LM3-burst phase raises the LM3 arrival range while keeping the other LM types at the normal range. The LM1/LM4-burst phase raises the LM1 and LM4 arrival ranges simultaneously. The LM3 burst isolates a skew toward one image service, whereas the LM1/LM4 burst combines a text-service surge with the most resource-intensive image service. This final phase therefore changes both request volume and service composition, creating a capacity-limited regime. Each burst persists for multiple control intervals, allowing the telemetry to capture sustained queueing and deployment responses under the new demand regime.

Figure~\ref{fig:bursty}(b) separates the methods by their ability to maintain a low latency--fairness objective after the workload mix changes. MA remains below both baselines in all three regimes, and the gap becomes more visible after the burst transitions. The contrast between MA and MAL isolates the value of coordinated local deployment. Both methods apply agentic traffic planning and prompt routing, but MAL pairs this routing layer with an independently reacting Lyapunov-based deployment algorithm, and its objective rises much more sharply under the LM3 burst. A comparison between MA and PerLLM gives a different pattern. PerLLM changes smoothly across the burst phases, which is consistent with its online bandit updates, but its objective remains higher than MA in both shifted regimes. This gap suggests that adapting the routing learner from delayed rewards is less effective when the deployment layer is optimized separately from the routing decisions. In the last phase, all methods degrade because the workload itself consumes more serving capacity, but MA still retains the lowest objective. These results show that MA preserves a lower objective across sustained workload shifts without re-training or re-initializing the agents.

To evaluate the proposed framework under an expanded MEC network, we add two worker VMs at the University of Bristol site on top of the original multi-site testbed.
The two added VMs follow the hardware sets used in Table~\ref{configuration}: one single-GPU VM matching VM3--VM6 and one dual-GPU VM matching VM7. The same agent logic and per-node workload-generation rule are then used in the expanded deployment, so the number of worker-side agents increases from $6$ to $8$ and the average request load increases from approximately $96$ to $128$ requests per slot. Table~\ref{tab:scalability} shows that MA maintains a comparable objective value in this larger setting. The Jain's index increases from $0.925$ to $0.932$, while latency increases from $96.7$\,s to $132.1$\,s under the higher aggregate load. This pattern suggests that the expanded GPU-capable service coverage helps balance LM service types, while the higher aggregate request volume and larger routing space still introduce additional queueing cost. This result demonstrates that the proposed agent interfaces remain effective when the worker-agent set and the request volume scale together. Adding MEC servers enlarges the planning and routing problem, but LM deployment remains distributed across the added nodes.

Overall, the results show that MA balances latency and fairness, remains robust under delayed feedback and workload shifts, and maintains its effectiveness as the number of workers increases. However, several limitations remain. At the implementation level, cloud-hosted foundation-model APIs introduce dependence on external service availability, response latency, pricing, and model-version changes. From the objective perspective, the fairness analysis is presently defined at the service-type level. Extending it to user-level or priority-aware objectives requires a richer request model. Regarding scale, the six-worker and eight-worker testbeds provide initial evidence for a larger deployment, while substantially larger networks remain to be evaluated. On the theoretical side, the stability analysis provides a conditional queue-drift interpretation. Formal guarantees of Tier-1 convergence, monotonic improvement, and global optimality remain open. Future work will investigate self-hosted agent execution, richer fairness models, larger deployments, and hybrid agentic-control formulations amenable to formal performance analysis. 

\vspace{-0.4cm}
\section{Conclusion}
\label{sec:conclusion}
This paper presents a hierarchical Agentic AI framework for optimizing latency-sensitive and fair inference of multi-modal LMs in mobile edge networks. By coordinating a Global Planning Agent with a Prompt Scheduling Agent and on-node LM Deployment Agents, we demonstrate that system-level objectives can be achieved through natural-language reasoning over runtime telemetry and historical memory.  Our framework integrates clean modular abstractions with a fully functional edge testbed, enabling end-to-end coordination across dynamic prompt workloads and resource-constrained edge nodes.
Extensive experiments show that MA reduces average end-to-end latency by 42.9\% and 21.4\% compared with PerLLM and MAL, respectively, while maintaining a competitive mean service-type Jain's index of 0.925. It also achieves an 80.6\% latency reduction relative to the full-activation baseline and demonstrates stronger adaptation performance under delayed and non-stationary feedback.
Looking ahead, future work will explore joint optimization of service latency and energy consumption, enabling sustainable LM deployment in resource-constrained environments. 

\bibliographystyle{IEEEtran}
\bibliography{references}

\end{document}